\documentclass[aps,prd,superscriptaddress,twocolumn,tightenlines]{revtex4}
\input epsf
\usepackage{amsmath,amssymb,subfigure}
\usepackage{graphicx}
\usepackage{epsfig}
\usepackage{color}
\usepackage{ulem}

\newcommand{\be}{\begin{equation}}
\newcommand{\ee}{\end{equation}}

\begin{document}

%%%%%%%%%%%%%%%%%%%%%%%%%%%%%%%%%%%%%%%%%%%%%%%%%%%%%%%%%%%%%%%%%%%%%%%%%%%%%%%%%%%%%%%%%%%%  

\title{Large Scale Structure as a Probe of Gravitational Slip}

\author{Scott F. Daniel\footnote{scott.f.daniel@dartmouth.edu}}
\affiliation{Department of Physics and Astronomy, Dartmouth College, 
Hanover, NH 03755 USA}
\author{Robert R. Caldwell}
\affiliation{Department of Physics and Astronomy, Dartmouth College, 
Hanover, NH 03755 USA}
\author{Asantha Cooray}
\affiliation{Department of Physics and Astronomy, University of California, 
Irvine, CA 92697 USA}
\author{Alessandro Melchiorri}
\affiliation{Physics Department and Sezione INFN, University of Rome,
``La Sapienza,'' P.le Aldo Moro 2, 00185 Rome, Italy}
 
\date{\today}

\begin{abstract}

A new time-dependent, scale-independent parameter, $\varpi$, is employed in a
phenomenological model of the deviation from General Relativity in which the Newtonian and
longitudinal gravitational potentials slip apart on cosmological scales as dark energy,
assumed to be arising from a new theory of gravitation, appears to dominate the universe. A
comparison is presented between $\varpi$  and other parameterized post-Friedmannian models
in the literature. The effect of $\varpi$ on the cosmic microwave background anisotropy
spectrum, the growth of large scale structure, the galaxy weak-lensing correlation
function, and cross-correlations of cosmic microwave background anisotropy with galaxy
clustering are illustrated. Cosmological models with conventional maximum likelihood
parameters are shown to find agreement with a narrow range of gravitational slip.

\end{abstract}

\maketitle

%%%%%%%%%%%%%%%%%%%%%%%%%%%%%%%%%%%%%%%%%%%%%%%%%%%%%%%%%%%%%%%%%%%%%%%%%%%%%%%%%%%%%%%%%%%%
\section{Introduction}

The quest for the source of the cosmic acceleration has lead to speculation that the proper
theory of gravitation departs from General Relativity (GR) on cosmological scales. That is,
gravitation may be well described using Einstein's theory within the solar system and the
environs of the galaxy, but a different theory is required on the scale of the Hubble
length (e.g. Ref.~\cite{Uzan:2006mf}). There are numerous examples of a theory capable of
producing a late-time acceleration, all of which introduce new gravitational degrees of
freedom, with wide-ranging implications for observable phenomena
\cite{Carroll:2004de,Nojiri:2006ri}. Given this possible abundance in new physics, it is
important to come up with tests that can distinguish between the effects of dark energy and
those of modified gravity. Though late-time accelerated cosmic expansion is the principal
indicator that a new ``dark'' physics is needed, it is not the only test such physics must
satisfy. A successful cosmology must also agree with measurements related to the growth of
perturbations as probed by the cosmic microwave background, clustering of large-scale
structure, and weak lensing deflections of light.

The concordance cosmology, $\Lambda$CDM within GR, gives very specific predictions for the
cosmic expansion and the evolution of inhomogeneities. It has been proposed to evaluate
alternative theories of gravitation by testing for violations of these predictions (e.g.
Ref.~\cite{Ishak:2005zs}). Efforts along these lines consist of fitting separate
$\Lambda$CDM parameters to tests of structure growth and cosmic expansion
\cite{Wang:2007fsa}, or constraining a phenomenological description of the linear growth
rate of density perturbations with galaxy clustering surveys
\cite{Huterer:2006mva,Linder:2007hg,DiPorto:2007ym,Nesseris:2007pa}. To within the stated
uncertainties, these current results are all consistent with $\Lambda$CDM. The absence of
any realistic theory  of the cosmological constant, however, has prompted many to question
whether gravity itself is the culprit.

An exhaustive study of departures from GR as an explanation for dark energy phenomena is
difficult, as Einstein's theory represents a mere island in a sea of possible
gravitational theories. Yet, a common feature within a broad range of such theories is a
decoupling of the perturbed Newtonian-gauge gravitational potentials $\phi$ and $\psi$,
defined by the perturbed Robertson-Walker line-element
\begin{equation}
\label{metric}
ds^2=a^2 \left[-\left(1+2\psi\right)d\tau^2
+\left(1-2\phi\right)d\vec{x}^2\right] \, ,
\end{equation}
using the notation and convention of Ma \& Bertschinger  \cite{Ma:1995ey}.  To give a
physical sense of these potentials, $\psi$ enters the Newtonian limit of the equation of
motion, $\ddot{\vec x} = -\vec\nabla\psi$, and $\phi$ enters the Poisson equation
$\nabla^2\phi=-4\pi G a^2\delta\rho$. 

Whereas GR predicts $\psi=\phi$ in the presence of non-relativistic matter, a {\it
gravitational slip}, defined as $\psi\neq\phi$, generically occurs in modified gravity
theories. This inequality means that the gravitational potential created by a galaxy
cluster is not the same potential responsible for the geodesic motion of the constituent
galaxies. For primordial cosmological perturbations, the potentials are not completely
free, however, as there exists a constraint equation, valid under general assumptions  in
the long-wavelength limit \cite{Bertschinger:2006aw}. Hence, a new relation between these
potentials is a launching point for investigations of cosmological manifestations of
modified gravitation. 

With a view towards testing modified gravitation against large-scale structure, the
relation between $\phi$ and $\psi$ has been examined for scalar-tensor theories
\cite{Schimd:2004nq}, generalized gravitational Lagrangians or $f(R)$ theories
\cite{Acquaviva:2004fv,Zhang:2005vt}, a tensor-vector-scalar model of gravity
\cite{Skordis:2005eu}, the Dvali-Gabadze-Porrati (DGP) model
\cite{Dvali:2000hr,Lue:2005ya,Song:2006jk}, and a Lorentz-invariance violating massive
gravity \cite{Bebronne:2007qh}. Phenomenological relations between the potentials have been
widely investigated
\cite{Lue:2003ky,Zhang:2007nk,Amendola:2007rr,Schmidt:2007vj,Amin:2007wi,Jain:2007yk,Bertschinger:2008zb}.
A general set of post-Friedmannian parameters has also been proposed \cite{Hu:2007pj},
which reproduces the evolution of $\phi$ and $\psi$ on both sub- and super-horizon scales
for DGP and $f(R)$ gravities; the relation between these parameters and cosmic acceleration
has been discussed in Ref.~\cite{Hu:2008zd}. Much of the analysis has been formal, with an
aim towards future tests. Few efforts have been made to place constraints using current
data. 

In this paper we explore the parameterized post-Friedmannian (PPF) description of
gravitation introduced in Ref.~\cite{Caldwell:2007cw} (hereafter CCM). CCM posit a modified
theory of gravitation that produces a $\Lambda$CDM-equivalent background with perturbations
such that $\psi=(1+\varpi)\phi$, where $\varpi$ is taken to be scale-independent. Here, we
correct an important error in CCM, and explore the impact on a wider range of cosmological
phenomena. In Sec.~\ref{paramsection} we review the model for the time-dependent quantity
$\varpi$ and the description in terms of cosmological parameters. We compare  the PPF
formalism to similar attempts at modifying GR in the literature.  In Sec.~\ref{pertsection}
we give the procedure for evolving cosmological perturbations and describe modifications to
the CMBfast software \cite{Seljak:1996is}.  In Sec.~\ref{cmbsection} we explore the
influence of $\varpi$ on the  cosmic microwave background (CMB) anisotropy, correcting
several errors in the numerical calculations described in CCM.   We show the effect of
$\varpi\ne0$ on the Wilkinson Microwave Anisotropy Probe (WMAP) \cite{wmapdata} best-fit
cosmology and its consistency with current data. Using WMAP best-fit cosmological
parameters, we  similarly demonstrate the effect of $\varpi\ne0$ on perturbation growth in
Sec.~\ref{growthsection}, weak lensing in Sec.~\ref{wlsection}, and the integrated
Sachs-Wolfe effect in Sec.~\ref{iswsection}. We attempt a synthesis of these results in
Sec.~\ref{secsynth}, and present a final discussion in Sec.~\ref{secdiscuss}.

%%%%%%%%%%%%%%%%%%%%%%%%%%%%%%%%%%%%%%%%%%%%%%%%%%%%%%%%%%%%%%%%%%%%%%%%%%%%%%%%%%%%%%%%%%%%
\section{Parameterization}
\label{paramsection}

Non-Einstein gravitation generically predicts a decoupling of the gravitational potentials,
$\phi \neq \psi$, such that the two potentials are independent functions of the four
space-time coordinates, $\tau$ and $\vec{x}$:
\begin{equation}
\psi(\tau,\vec x) =\left[1+\varpi(\tau,\, \vec x)\right] \times \phi(\tau,\vec x) \, .
\label{varpidef}
\end{equation}
The model proposed in CCM attempts to link the departure from GR with the growth of  an
effective dark energy relative to normal matter and radiation, whereby it was motivated
that
\begin{eqnarray}
\varpi(\tau,\, \vec x) &=& 
\varpi_0 \rho_{\rm DE}(\tau,\,\vec x) / \rho_{\rm m}(\tau,\,\vec x), \cr\cr
&\approx & \varpi_0 \frac{\bar \rho_{\rm DE}(\tau)}{\bar\rho_{\rm m}(\tau)}.
\end{eqnarray}
The second line is obtained by expanding to first order in perturbations on the homogeneous
background, indicated by the overbar. Because $\varpi$ already multiplies first order
perturbation quantities, the spatial dependence enters at second order and is neglected
here. Truly, any theory of gravitational slip must be scale dependent in order to both
satisfy  solar system constraints at small length scales and allow for novel cosmological
phenomena at extra-galactic length scales.  For the purposes of our investigation, focusing
on cosmology, $\varpi$ is considered to be homogeneous.

The background evolution is taken to be identical to $\Lambda$CDM, in which case
\begin{equation}
\varpi = \varpi_0\frac{\Omega_{\Lambda}}{\Omega_m} (1+z)^{-3}.
\label{varpievolution}
\end{equation}
Hence, the gravitational slip remains negligible until late times, when the cosmic
acceleration becomes apparent. Our PPF cosmological model is described by the standard set
of $\Lambda$CDM parameters, plus $\varpi_0$. Informally, this model may be referred to as
``$\varpi\Lambda$CDM''. While there is no {\it a priori} reason that new gravitational
physics should behave this way, this model provides a simple means to test for indications
of new gravitational physics. If cosmological data are found to be favoring a non-zero
value for $\varpi$, an advanced theory of gravity beyond GR will be required to explain its value and redshift evolution.

Our parameterization of the relationship between the gravitational potentials can be
compared to modifications of GR in the recent literature. Bertschinger \& Zukin
\cite{Bertschinger:2008zb} (hereafter BZ) adopt a notation with $\psi=\Phi_{\rm BZ}$ and
$\phi=\Psi_{\rm BZ}$ and parameterize
\begin{equation}
\label{BZeqn}
\Psi_{\rm BZ}({\bf k},t) = \gamma_{\rm BZ}(a) \Phi_{\rm BZ}({\bf k},t) + ...\, ,
\end{equation}
with $\gamma_{\rm BZ}(a)=1+\beta a^s$ where $\beta$ and $s>0$ are model parameters. In our
language (\ref{BZeqn}) takes the form $\phi=\gamma_{\rm BZ}(a)\psi$, with $\gamma_{\rm
BZ}(a)=1/(1+\varpi(a))$ leading to
\begin{equation}
\varpi(a)=-\beta a^s (1+\beta a^s)^{-1} \, .
\label{eqn:BZ}
\end{equation}
In the case $s=3$ and the limit of small $\beta$, the parametrization of BZ agrees with our
model (\ref{varpievolution}) for $\beta=-\varpi_0 \Omega_{\Lambda}/\Omega_m$. As we discuss
later, CMB anisotropy spectra obtained using the parameterization of
equation~(\ref{eqn:BZ}) in our code are fully consistent with the results of BZ.  The
differences in the CMB anisotropy spectra reported in BZ and CCM are due to a numerical
error in the implementation of CCM, which we correct as discussed below. 

Hu \cite{Hu:2008zd} presents another parameterization with
\begin{equation}
\frac{\Phi_{\rm Hu}+\Psi_{\rm Hu}}{2}=g(k,a)\left[\frac{\Phi_{\rm Hu}-\Psi_{\rm Hu}}{2}\right] + ... \, ,
\end{equation}
with $g(k,a)=g_{\rm SH}(a)/(1+c_g^2k_H^2)$ and $g_{\rm SH}(a)=g_0\sqrt{\rho_{\rm de}
\Omega_{\rm tot}/\rho_{\rm tot} \Omega_{\rm de}}$. In our notation the perturbations are
now $\phi=-\Phi_{\rm Hu}$ and $\psi=\Psi_{\rm Hu}$. Ignoring the momentum dependence, which
only dampens the post-Friedmannian modification below a certain scale, the correspondence
is found to be
\begin{equation}
\label{HUeqn}
\varpi(a)=\frac{-2g(a)}{1+g(a)} \, .
\end{equation}
Given the difference in redshift dependence between equations (\ref{varpievolution}) and
(\ref{HUeqn}) for the $g(a)$ used in Ref.~\cite{Hu:2008zd}, a simple
relation between $\varpi_0$ and $g_0$ that is accurate at redshifts other than $z=0$
is not evident. Thus, a detailed comparison is not pursued.

Jain \& Zhang \cite{Jain:2007yk} present a post-Friedmannian parameterization with
$\phi=\eta(k,t)\psi$. Ignoring the momentum dependence, $\eta$ is related to $\varpi(a)$
through $\varpi(a)=\eta(a)^{-1}-1$. Since they do not present numerical results on the
modification imposed by their parameterization to CMB anisotropies and other large-scale
structure observables, a comparison with their work is also not attempted here.
 
In addition to modifications of gravitation that lead to $\phi\neq \psi$, constraints on
possible departures from the Newtonian inverse-square law at cosmological distance scales
have also been considered \cite{Sealfon:2004gz,Dore:2007jh}. It is not certain that such
modifications satisfy the consistency relation for cosmological perturbations described in
Ref.~\cite{Bertschinger:2006aw}. In any event, our study differs from such works in
that no scale-dependent corrections to gravitation are imposed.

%%%%%%%%%%%%%%%%%%%%%%%%%%%%%%%%%%%%%%%%%%%%%%%%%%%%%%%%%%%%%%%%%%%%%%%%%%%%%%%%%%%%%%%%%%%%
\section{Implementation of PPF Model}
\label{pertsection}

An implementation of the parametrized relationship between the potentials requires a
consistent  method of treating perturbations in the modified theory. To begin, the
phenomenological relation between the gravitational potentials is defined in the
conformal-Newtonian (longitudinal) gauge. In practice, however, we find it most convenient
to pursue perturbation evolution in the synchronous gauge. Using the notation of
Ref.~\cite{Ma:1995ey}, the metric perturbation variables in the two gauges are related as
\begin{eqnarray}
\psi&=&\frac{1}{2k^2}\left[\ddot{h}+6\ddot{\eta}+\mathcal{H}\left(\dot{h}+6\dot{\eta}\right)\right]\\
\phi&=&\eta-\frac{1}{2k^2}\mathcal{H}\left(\dot{h}+6\dot{\eta}\right)\, ,
\label{phipsi}
\end{eqnarray}
where the dot indicates the derivative with respect to conformal time. In the standard GR
case ($\varpi=0$), the perturbed Einstein equations, 
\begin{eqnarray}
k^2\eta-\frac{1}{2}\mathcal{H}\dot{h}&=&4\pi Ga^2\delta T^0_0\label{zerozero}\\
k^2\dot{\eta}&=&4\pi G a^2 (\bar{\rho}+\bar{p}) \theta\label{zeroi}\\
\ddot{h}+2\mathcal{H}\dot{h}-2k^2\eta&=&-8\pi G a^2\delta T^i_i \, ,\label{ii}
\end{eqnarray}
are used to evolve the metric variables, where $(\bar{\rho}+\bar{p})\theta\equiv i
k^j\delta T^0_j$ (for greater detail, see \cite{Ma:1995ey}).  It is standard practice to
use the latter two equations (\ref{zeroi},\,\ref{ii}) for evolution, and apply the first
equation (\ref{zerozero}) as a constraint.

We presume a theory of modified gravitation in which the stress-energy tensor of matter and
radiation is conserved, equation (\ref{zeroi}) remains valid, but equations
(\ref{zerozero},\,\ref{ii}) are invalid. If the perturbed Einstein equations were assumed
to remain valid, then the gravitational slip would necessarily imply the existence of new
energy density and pressure perturbations which are comoving with the baryonic and dark
matter density perturbations. Furthermore, the non-zero gravitational slip introduces a
modification to the perturbed, off-diagonal space-space Einstein equation  
\begin{equation}
\dot{\alpha}=-(2+\varpi)\mathcal{H}\alpha+(1+\varpi)\eta  
 -12\pi G a^2(\bar{\rho}+\bar{p})\sigma/k^2 \label{alphadot}
\end{equation}
where $\alpha \equiv (\dot h + 6 \dot\eta)/2 k^2$. The factor of $1/k^2$ in the last term
on the right hand side of (\ref{alphadot}) corrects a typographical error in CCM
equation~(8). Note that the shear due to matter and radiation
is negligible at late times. This leaves $\dot{\alpha} =-(2+\varpi)\mathcal{H}\alpha
+(1+\varpi)\eta$ which is simply a restatement of (\ref{phipsi}) in synchronous gauge. In
order to maintain continuity from early times, when GR is valid, to late times, when the
gravitational slip becomes significant, equation (\ref{alphadot}) is used. In turn,
\begin{eqnarray}
\label{alphaddot}
\ddot{\alpha}&=&\dot{\eta}(1+\varpi)-(2+\varpi)(\dot{\mathcal{H}}\alpha+
\mathcal{H}\dot{\alpha})+\dot{\varpi}(\eta-\mathcal{H}\alpha)\cr
\qquad&& -\frac{d}{d\tau}\left(12\pi G a^2(\bar{\rho}+\bar{p})\frac{\sigma}{k^2}\right).
\end{eqnarray}
Now, instead of evolving equations (\ref{zeroi},\,\ref{ii})  with constraint
(\ref{zerozero}), equations (\ref{zeroi},\,\ref{alphadot}) are evolved with the constraint
\begin{equation}
\label{hdot}
\dot{h}=2k^2\alpha-\frac{24\pi Ga^2}{k^2}(\bar{\rho}+\bar{p})\theta
\end{equation}
derived from $\alpha$ and equation (\ref{zeroi}). Thus, the time evolution of $\eta,\,h$
and $\alpha,\,\dot\alpha$ are fully specified in a modified gravity model.

The above set of equations, corresponding to the recipe R1 of CCM, has been implemented in a
modified version of CMBfast \cite{Seljak:1996is}. As shown in Appendix~\ref{zetasection},
this prescription satisfies the consistency condition that was first derived in
Ref.~\cite{Bertschinger:2006aw} (equation~2 of BZ). This consistency suggests that our
implementation, introduced by CCM, is not fundamentally different from that employed by BZ.
Despite the reasons suggested in BZ for differences between the two approaches, consistent
results are obtained when the two post-Friedmannian parameterizations are matched.

Note that when presenting results here, the error in the numerical software used by CCM,
involving a minus sign error in the $\dot{\varpi}$ term of equation~(\ref{alphaddot}), has
been corrected. Having rechecked the derivations and software, the revised results for the
effect of $\varpi$ on the CMB anisotropy power spectrum are described below.

%%%%%%%%%%%%%%%%%%%%%%%%%%%%%%%%%%%%%%%%%%%%%%%%%%%%%%%%%%%%%%%%%%%%%%%%%%%%%%%%%%%%%%%%%%%%
\section{Effects on CMB Anisotropy}
\label{cmbsection}

\begin{figure}[t]
\includegraphics[scale=0.35]{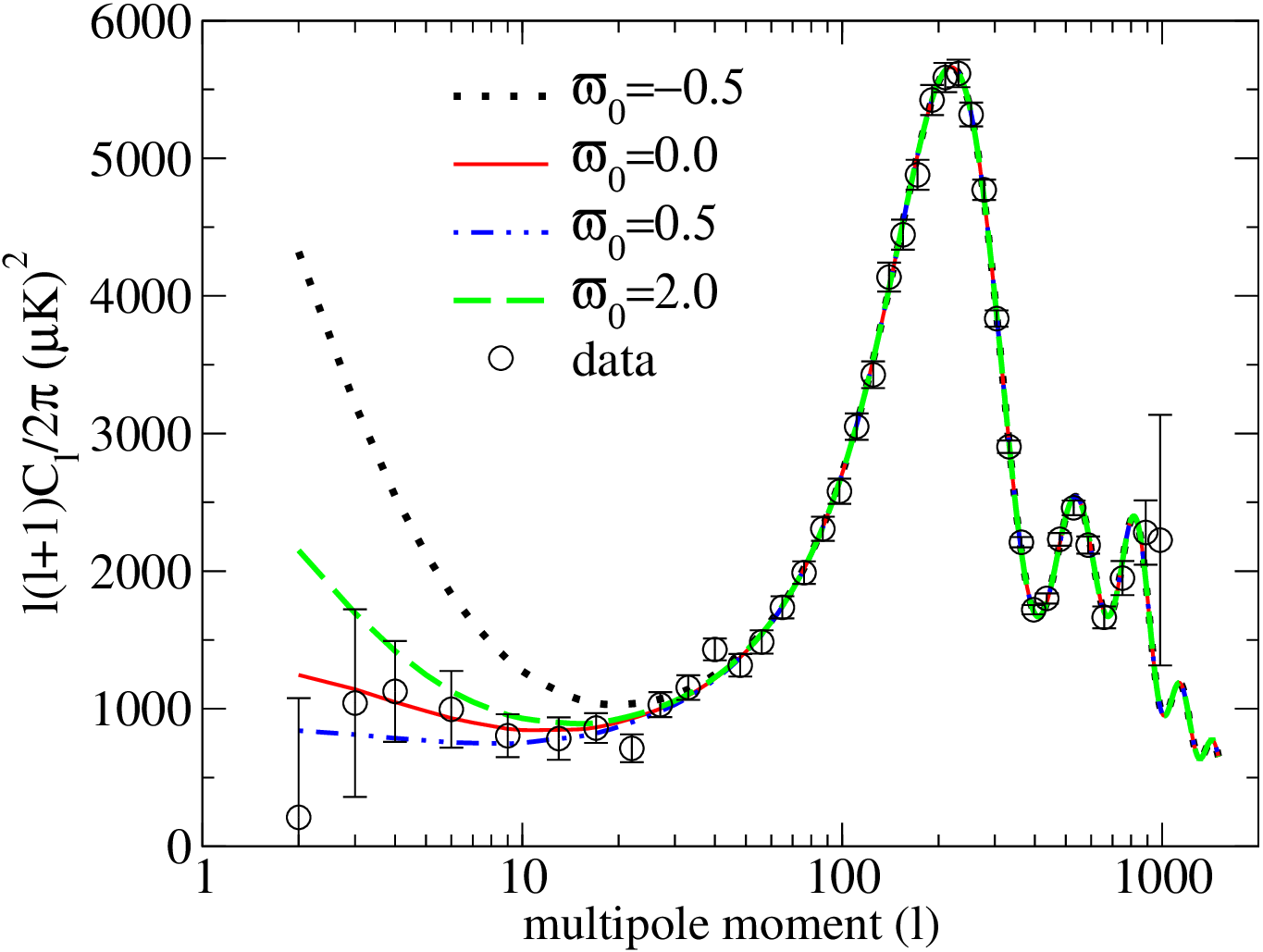}
\caption{CMB anisotropy power spectra for different $\varpi_0$ cosmologies are shown with
the binned WMAP 3-year data. All differences are localized to the low multipole moments:
the spectra are normalized so that the higher multipole moments for all models are
identical to the case of $\varpi_0=0$, corresponding to the WMAP3 ML model.}
\label{clwithdata}
\end{figure}

The predicted CMB anisotropy spectra are compared with the WMAP 3-year data, and a
likelihood function for the modified gravity variable $\varpi_0$ is constructed, using the
likelihood code supplied by the WMAP team \cite{wmapdata}. To avoid duplicating a lengthy
search of parameter space, the background cosmology is set to the 3-year WMAP maximum
likelihood cosmology (hereafter WMAP3 ML: $\Omega_{\rm b}=0.0414$, $\Omega_{\rm c}=0.196$,
$\Omega_\Lambda=0.7626$, $h=0.732$, $\tau=0.091$, $n_s=0.954$), and for comparison to
the 3-year WMAP plus SN Gold maximum likelihood cosmology (hereafter WMAP3+SNGold ML: 
$\Omega_{\rm b}=0.0454, \Omega_{\rm c}=0.2306, \Omega_\Lambda = 0.724, h=0.701, \tau=0.079,
n_s=0,946$) \cite{Spergel:2006hy}. Figure \ref{clwithdata} plots the multipole moments for
several simulations with  different values of $\varpi_0$ in the WMAP3 ML cosmology.
Normalizing the spectra to the observed amplitude of the acoustic peaks at $\ell \gtrsim
100$, we see that only the lowest multipole moments, $\ell \lesssim 30$, are affected by
$\varpi_0$.

\begin{figure}[t]
\includegraphics[scale=0.35]{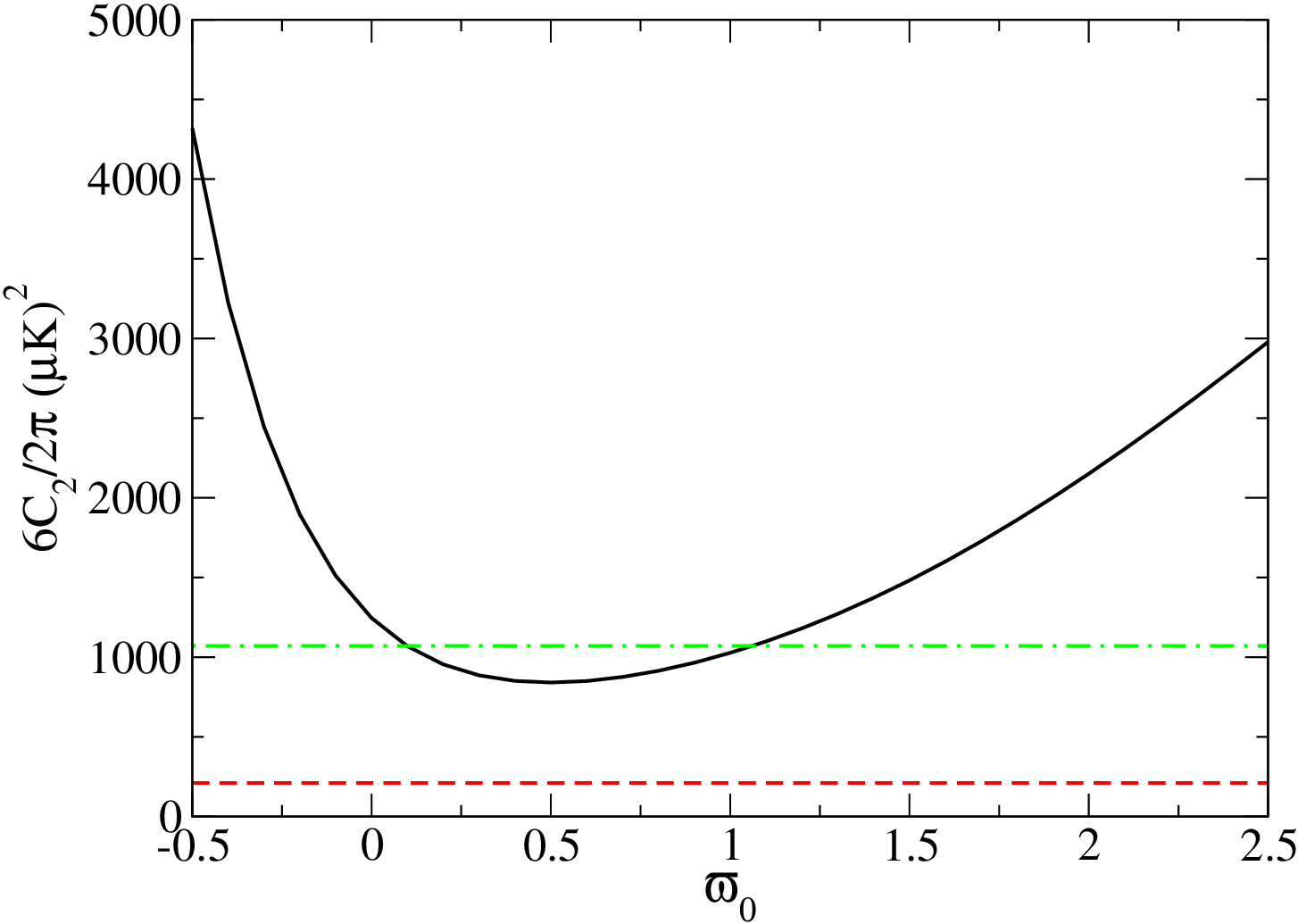}
\caption{The predicted CMB temperature anisotropy quadrupole power    is shown as a
function of $\varpi_0$ (solid curve). All other parameters are set to the WMAP3 ML cosmology.
The central value and 1$\sigma$ upper bound of the WMAP 3 year data, $6 C_2/2\pi=211\pm860$
\cite{wmapdata}, are shown by the dashed and dot-dashed curves.}
\label{quadfig}%
\end{figure}
 
Furthermore, we observe that the large-angle anisotropy power grows as $|\varpi_0-0.5|$
deviates from zero. For a closer look, the quadrupole moment is shown as a function of
$\varpi_0$ in Figure~\ref{quadfig}. This quadratic behavior can be understood as follows.
The integrated Sachs-Wolfe effect contribution to the temperature anisotropy  depends on
$\dot\phi+\dot\psi$, which varies with $\varpi$ as
\begin{equation}
\dot\phi + \dot\psi = \left(3 \varpi + (\Gamma-1)(2+\varpi)\right) {\cal H}\phi
\label{iswanalysis}
\end{equation}
where $\Gamma \equiv 1+d\ln \phi/d \ln a$. Numerical results show that $\Gamma$ decreases
monotonically, becoming negative with increasing $\varpi_0$. Upon squaring the above
quantity, it is evident that $|\varpi| \gg 1$ will lead to strong anisotropy in the low
multipoles, which are sourced at late times. Note that a similar trend, a local minimum of
large-angle anisotropy power as a function of the post-Friedmannian parameter, is seen in
Hu's model, in Figure 1 of Ref.~\cite{Hu:2008zd}. BZ should see the same trend, if a wider
range of $\beta$ is searched than in Figure 5 of Ref.\cite{Bertschinger:2008zb}.

\begin{figure}[ht]
\includegraphics[scale=0.35]{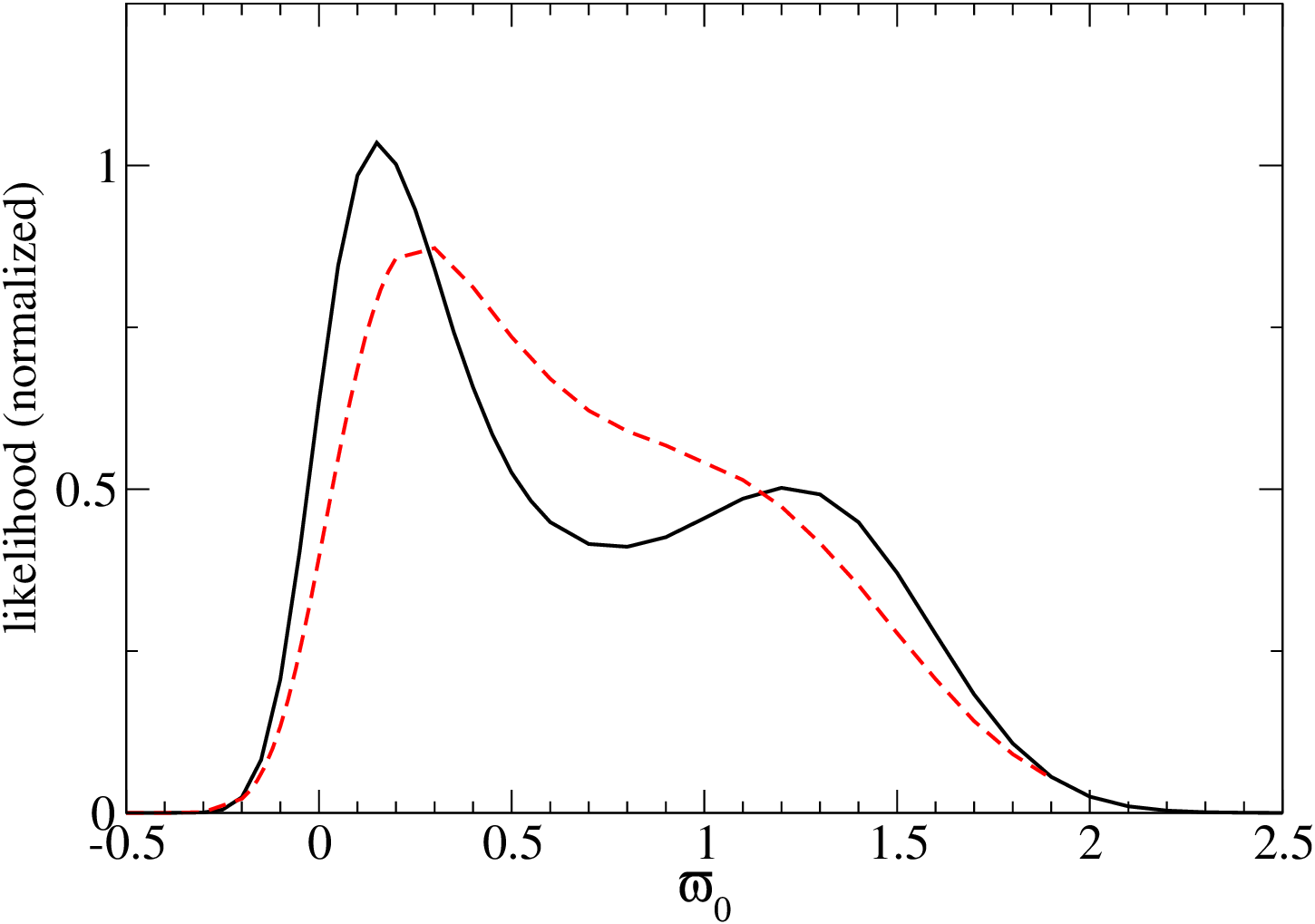}
\caption{The likelihood of $\varpi_0$ due to the CMB, calculated using the November 2006
version of the WMAP likelihood code \cite{wmapdata}, is shown.  All other cosmological
parameters are set to the WMAP3 ML (solid curve) or the WMAP3+SNGold ML (dashed curve)
model. The results include the TT and TE spectra. The results are consistent with
$\varpi_0=0$ ($\Lambda$CDM), but favor positive values of $\varpi_0$. The locations of the
primary (left) and secondary (right) likelihood peaks correspond to the range of $\varpi_0$
for which the predicted quadrupole lies within the 2$\sigma$ range of WMAP. }
\label{wmaplikelihood}%
\end{figure}

The local minimum in the quadrupole amplitude helps explain the double-peaked structure of
the WMAP likelihood function, shown in Figure \ref{wmaplikelihood}. The likelihood is
suppressed at large values of $|\varpi_0-0.5|$, due to the excessive large angle anisotropy
power. And while the low quadrupole moment reported by WMAP has been interpreted to be
indicative of new physics, in fact the WMAP likelihood does not necessarily reward an
anisotropy spectrum that smoothly leads to low power at low $\ell$. The anisotropy spectrum
with lowest quadrupole, at $\varpi_0 = 0.5$, is suppressed relative to $\varpi_0=0,\,1.5$
for the WMAP3 ML parameters,  as shown in Figure~\ref{wmaplikelihood}.  In the case of the
WMAP3+SNGold ML model, which has a slightly higher matter density, the primary  peak in the
likelihood distribution is shifted towards a slightly higher value of $\varpi_0$.  To
understand this behavior, note that in the case $\varpi=0$, the strength of the ISW
contribution depends on $\Omega_m$, with $\Gamma\approx(\Omega_m[a])^{0.55}$. An increase
in $\Omega_m$ thereby diminishes the ISW contribution. Because the WMAP CMB data appears to
prefer a moderate ISW contribution, the low quadrupole notwithstanding, then the data
should prefer a low matter density at $\varpi_0=0$, as seen in
Figure~\ref{wmaplikelihood}.  

\begin{figure}[ht]
\includegraphics[scale=0.35]{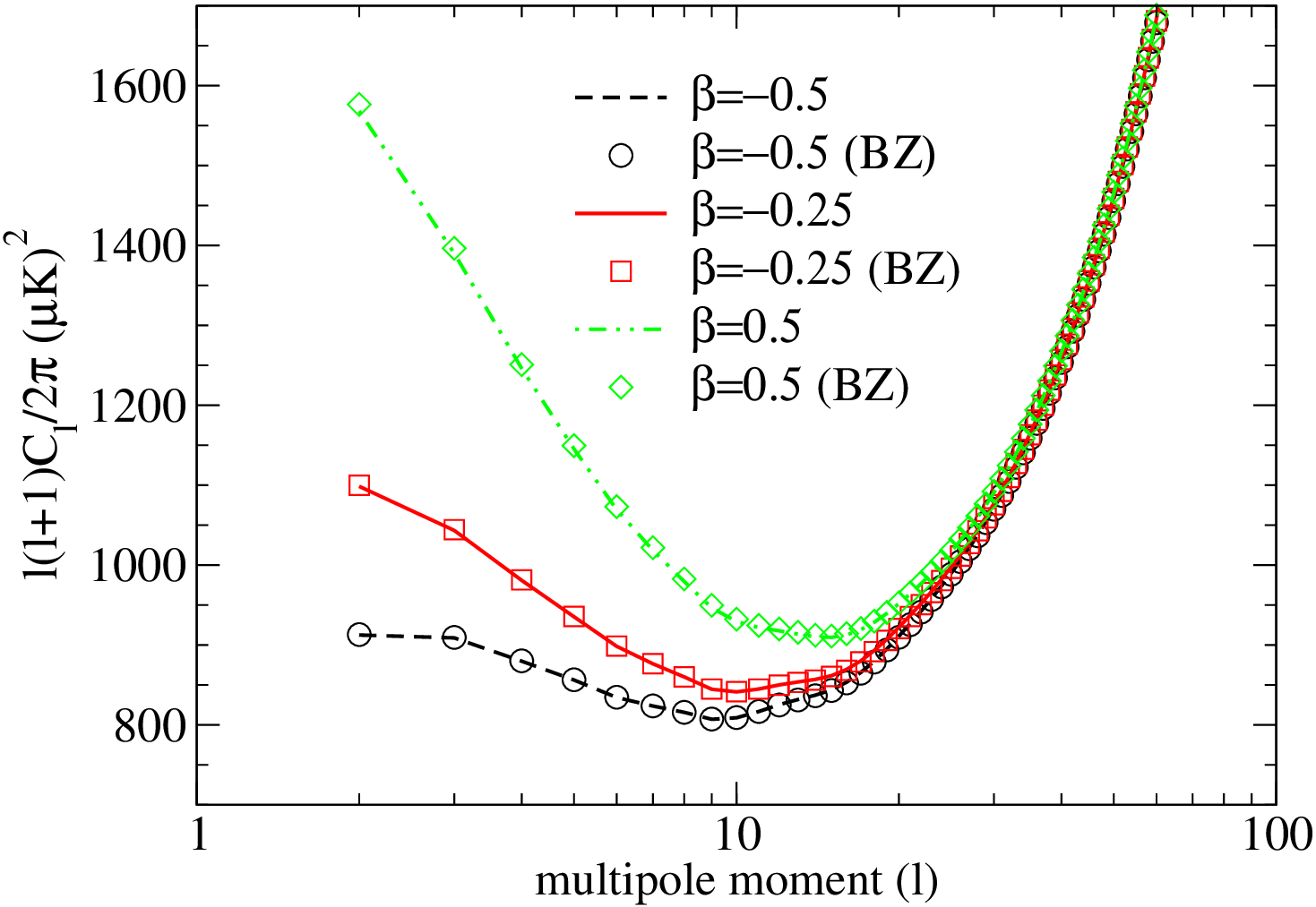}
\caption{The equivalence of alternative techniques for evolving cosmological  perturbations
under PPF gravitation is illustrated by the effect on CMB anisotropy spectra. Lines show
the spectra produced following our procedure (equations \ref{alphadot}-\ref{hdot}) with the
parameterization (\ref{varpievolution}) replaced by that proposed in BZ, $\varpi(a)=-\beta
a^s/(1+\beta a^s)$ with $s=3$. Symbols indicate the spectra produced following BZ's
procedure (equations \ref{zetaeqn}-\ref{alphadotbz}).  The agreement between the techniques
is exact.}
\label{bertschingerClvarpi}%
\end{figure} 

\begin{figure}[h]
\includegraphics[scale=0.35]{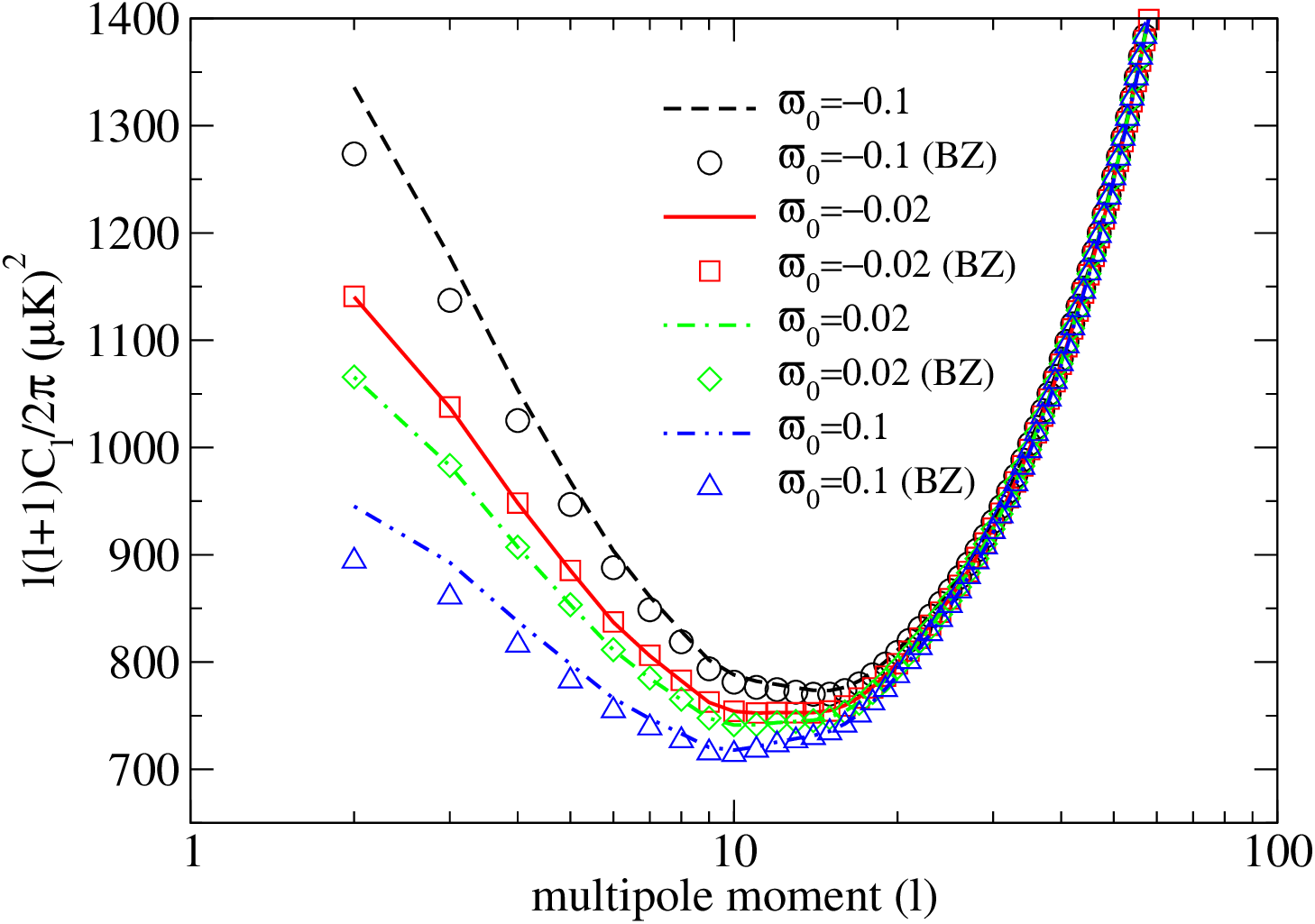}
\caption{The effect of the parameterization of gravitational slip on CMB anisotropy spectra
is shown. Lines show the spectra produced using our parameterization,  equation
(\ref{varpievolution}); symbols indicate spectra produced using the BZ's parameterization,
%$\varpi(a)=-\beta a^s/(1+\beta a^s)$ 
$\phi/\psi=1+\beta a^s$ with $s=3$. To compare, the parameter $\beta$ is set
to $\beta=-\varpi_0\Omega_\Lambda/\Omega_m$. The agreement is excellent in the limit of
small $\beta$.} 
\label{bertschingerClbeta}%
\end{figure}

Figures \ref{bertschingerClvarpi} and \ref{bertschingerClbeta} compare the results of BZ
with our PPF implementation. To recreate BZ's results, a second CMBfast
code was modified to evolve according to GR until $z=30$.  At that point, the parameter 
\begin{equation}
\label{zetaeqn}
\zeta=\frac{2}{3}\frac{\frac{\dot{\phi}}{\mathcal{H}}+\phi}{1+w}+\phi
\end{equation}
is calculated,
where $w$ is the background equation of state (assuming that only matter and $\Lambda$
contribute). Next, $\zeta$ is set to remain constant for the rest of the calculation,
whereupon the evolution is determined by the equations
\begin{eqnarray}
\dot{\phi}&=&(\zeta-\phi)\left(\mathcal{H}-\frac{\dot{\mathcal{H}}}{\mathcal{H}}\right)
-\mathcal{H}\psi
\label{phidotbz} \\
\dot{\alpha} &=& \psi-\mathcal{H}\alpha
\label{alphadotbz}
\end{eqnarray}
and $\psi=\phi/\gamma_{\rm BZ}$, $\gamma_{\rm BZ} = 1+\beta a^s$, $\eta =
\phi+\mathcal{H}\alpha$. Here, (\ref{phidotbz}) comes from equation (2) of BZ and
(\ref{alphadotbz}) comes from the definition of $\alpha$.  This second code uses none of
the Einstein equations.  The results, shown in Figures~\ref{bertschingerClvarpi} and
\ref{bertschingerClbeta}, are equivalent to our results produced using equation
(\ref{varpievolution}) if  the appropriate relationship between $\varpi$ and $\gamma_{\rm
BZ}$ is imposed. Appendix \ref{zetasection} demonstrates this equivalence analytically.

%%%%%%%%%%%%%%%%%%%%%%%%%%%%%%%%%%%%%%%%%%%%%%%%%%%%%%%%%%%%%%%%%%%%%%%%%%%%%%%%%%%%%%%%%%%%
\section{The Growth of Structure}
\label{growthsection}

The growth of density perturbations in baryonic and dark matter is affected by
gravitational slip, as the potential produced by an overdensity no longer matches the
potential responsible for gravitational acceleration. In GR, equations (\ref{zerozero}) and
(\ref{ii}) combine with the fluid conservation law (equation 29a in Ref.~\cite{Ma:1995ey}),
\begin{equation}
\label{deltadot}
\dot{\delta}=-(1+w)\left(\theta+\frac{\dot{h}}{2}\right)-
3\mathcal{H}\left(\frac{\delta{p}}{\delta{\rho}}-w\right)\delta,
\end{equation}
to give the equation for the growth of non-relativistic, pressureless perturbations:
\begin{equation}
\ddot \delta + {\cal H}\dot\delta =4 \pi G a^2\delta\rho .
\end{equation}
In the case $\varpi\neq 0$, an {\it independent} differential equation for the matter
density contrast is not readily available as equations (\ref{zerozero},\,\ref{ii}) have been
eliminated. CCM try to circumvent this obstacle, using (\ref{zeroi},\,\ref{alphadot}) and
$\alpha$, together with the assumption of negligible shear to arrive at
\begin{equation}
\ddot{\delta}+\mathcal{H}\dot{\delta}=k^2(1+\varpi)(\mathcal{H}\alpha-\eta).
\end{equation}
Unfortunately, that is as far as CCM can go without postulating another perturbed fluid
$\delta\rho_{\text{DE}}$ to account for the theory's departure from (\ref{zerozero}). 
However, there is no barrier to evolving $\delta$ -- it is simply obtained numerically
from (\ref{deltadot}). 
%\sout{A
%similar solution is attempted in Ref.~\cite{Jain:2007yk} wherein an effective, modified
%gravitational energy density $\rho_\text{MG}$ is posited (see their equations (11), (12),
%and (15)), which obeys the Poisson-like equation $
%k^2(\psi+\phi)=-8\pi \tilde{G}_\text{eff}(k,t)\bar{\rho}_\text{MG}a^2\delta_\text{MG}$.
%This still requires solving for the evolution of
%$\psi$ and $\phi$ to find  $\delta_\text{MG}$ must be found numerically.}

\begin{figure}[ht]
\includegraphics[scale=0.35]{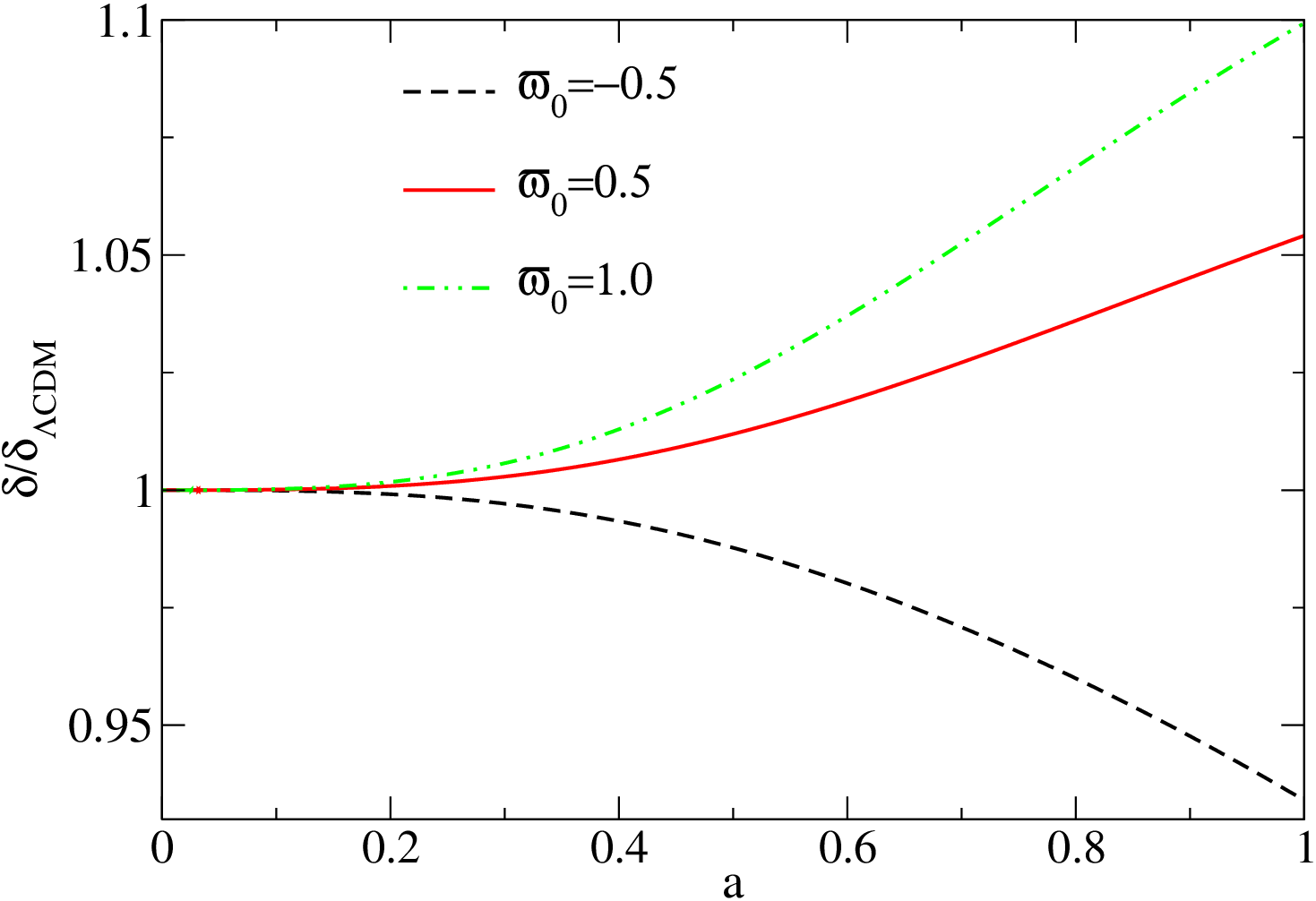}
\caption{The matter density contrast as a function of scale factor is shown for different
values of $\varpi_0$. The density contrast is normalized to a $\varpi_0=0$ model. All
models use WMAP3 ML parameters.}
\label{deltaratio} 
\end{figure}

\begin{figure}[ht]
\includegraphics[scale=0.35]{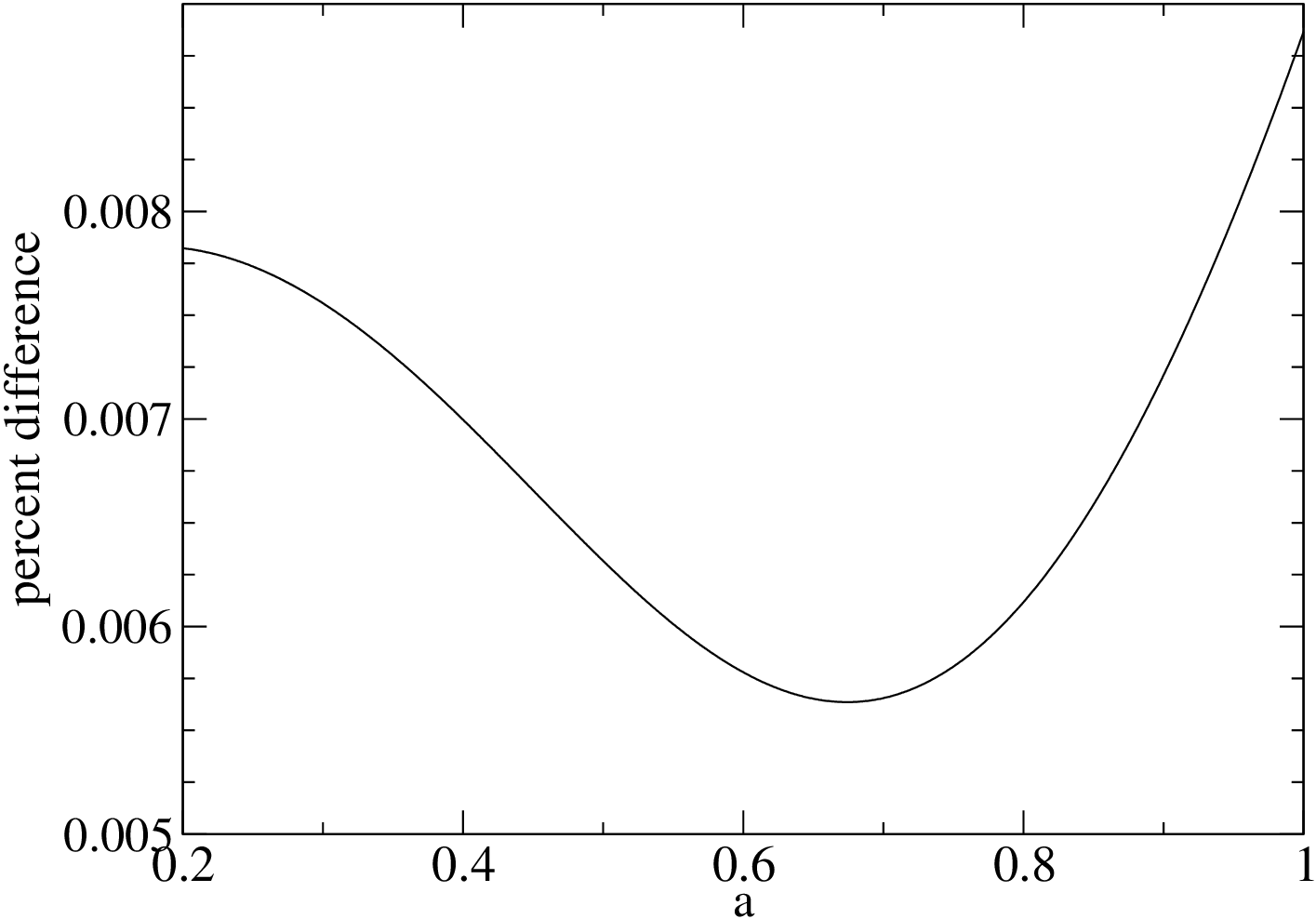}
\caption{The goodness of fit of the mapping of the growth rate between $\varpi\Lambda$CDM
and $\Lambda$CDM models, is illustrated. The percent difference in the density contrast for
a model with $\varpi_0=0.4$, $\Omega_m = 0.3$, and a $\Lambda$CDM model with
$\Omega_m\,=0.35$. These cosmologies are predicted to be equivalent by  equation
(\ref{omegamfit}).}
\label{deltatolcdm}
\end{figure}

The modified Boltzmann code CMBfast calculates $\delta$ as a function of time in the
process of calculating the anisotropy spectrum.  Figure~\ref{deltaratio} illustrates the
effect of $\varpi_0\ne 0$ on the growth of long wavelength matter density perturbations.  
This effect is monotonic with $\varpi_0>0$ enhancing the growth at late times and
$\varpi_0<0$ suppressing it.   By comparing the rate of growth for models with different
values of $\varpi_0$ and  $\Omega_m$, an approximate mapping for the growth of $\delta$
from $\varpi\Lambda$CDM to $\Lambda$CDM models has been obtained. Specifically,  to a good
approximation, the growth of structure in a $\varpi_0\ne 0$ cosmology is equivalent to a
$\Lambda$CDM cosmology with 
\begin{equation}
\label{omegamfit}
\Omega_{m}|_{\text{$\Lambda$CDM}}=\Omega_{m}|_{\varpi\Lambda{\rm CDM}}+0.13\varpi_0
\end{equation}
In the parameter ranges $-0.5\leq\varpi_0\leq0.5$ and
$0.2\leq\Omega_{m}|_{\varpi\Lambda{\rm CDM}}\leq0.5$, this fit is good to the $3\%$ level
for $a\geq0.2$. (See Figure \ref{deltatolcdm}). Note that $\varpi_0\ne 0$ does not change
the shape of the (linear) matter power spectrum.  

\begin{figure}[ht]
\includegraphics[scale=0.35]{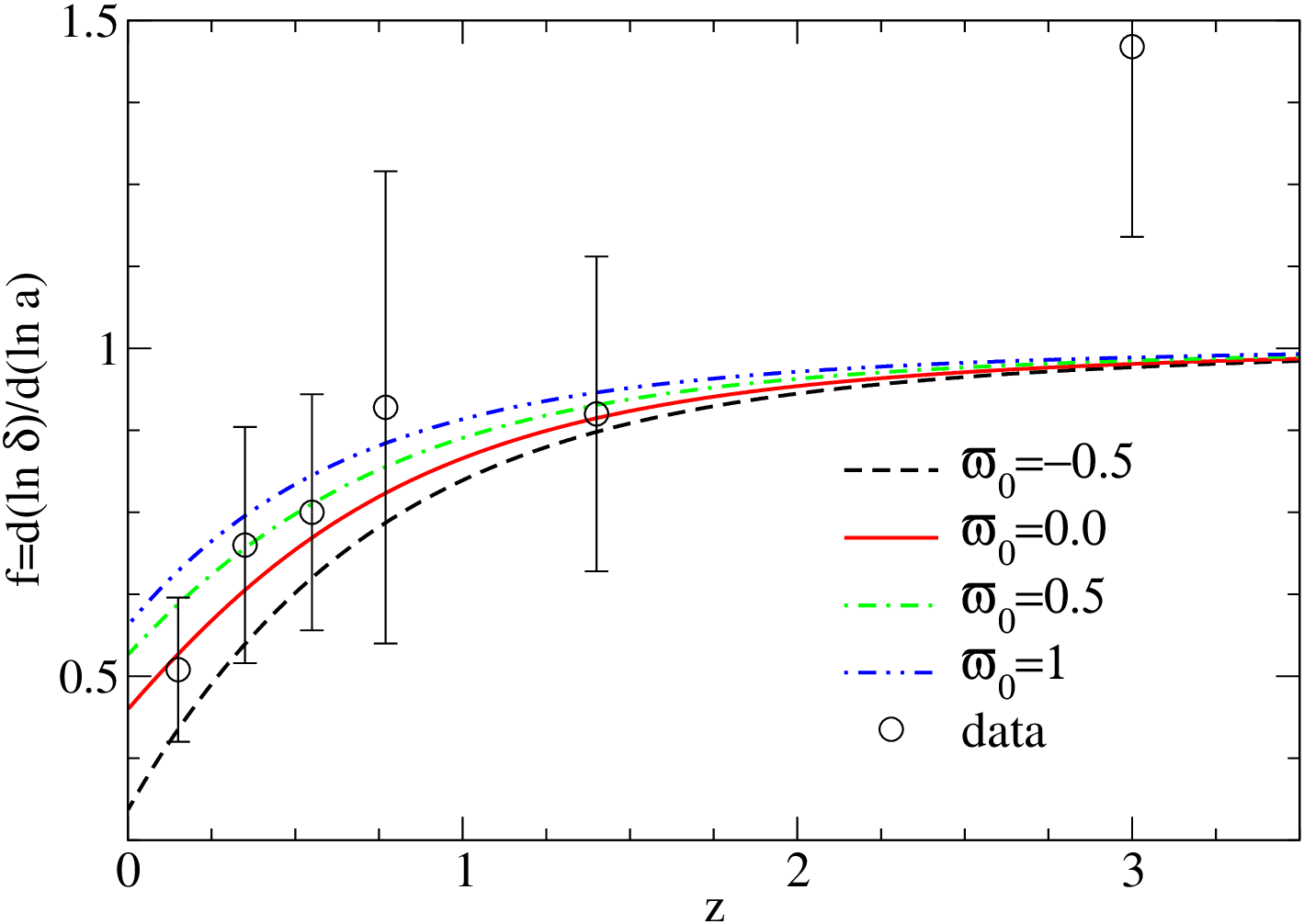}
\caption{The dimensionless linear growth rate $f$ plotted as a function of $z$ for
different $\varpi_0$ cosmologies with the WMAP3 ML background. The data point at $z=0.77$ is
due to Ref.~\cite{Guzzo:2008}. The remaining, hybrid set of data points are taken from
Table 1 of Ref.~\cite{Nesseris:2007pa}, and are based on estimates of large-scale structure
power spectrum growth at different redshifts.}
\label{NPplot}%
\end{figure}

Figure \ref{NPplot} plots the predicted dimensionless linear growth rate,
$f=d\ln\delta/d\ln a$, against a compilation of recent data. In the future, this may
provide an alternative test of modified gravitation.  Unfortunately, the current
uncertainties in $f$ are too large to provide a meaningful constraint on $\varpi_0$.

%%%%%%%%%%%%%%%%%%%%%%%%%%%%%%%%%%%%%%%%%%%%%%%%%%%%%%%%%%%%%%%%%%%%%%%%%%%%%%%%%%%%%%%%%%%%
\section{Weak Lensing}
\label{wlsection}

Gravitational lensing phenomena depends directly on the sum of the two gravitational
potentials, and is therefore an excellent probe of gravitational slip. In the cosmological
setting, measurements of weak lensing of the pattern of galaxy clustering can be used to
constrain $\varpi_0$. Proceeding, the E-mode weak lensing convergence correlation function,
$\xi_E$ \cite{Bartelmann:1999yn}, is calculated, which requires the convergence power
spectrum, given by \cite{lensing1,lensing2,lensing3,lensing4}
\begin{eqnarray}
P_\kappa(l)&=& l^4 \int_0^{\chi_\text{h}}d\chi\frac{n^2(\chi)}{r^4(\chi)} 
P_{\phi+\psi}^{\rm nonlin}\left(k=\frac{l}{r(\chi)},\chi\right)
\label{convergencespectrum}\\
n(\chi)&=&\int_\chi^{\chi_\text{h}}d\chi^\prime 
p(\chi^\prime)\frac{r(\chi^\prime-\chi)}{r(\chi^\prime)} 
\end{eqnarray}
where $\chi$ is the comoving radial distance with $\chi_\text{h}$ is the horizon distance, 
$r(\chi)$ is the comoving angular distance ($\chi$ in the case of a flat universe), $p(\chi)$ is the probability distribution of
lensed sources, and $0$ refers to the present epoch.
 
\begin{figure}[ht]
\includegraphics[scale=0.35]{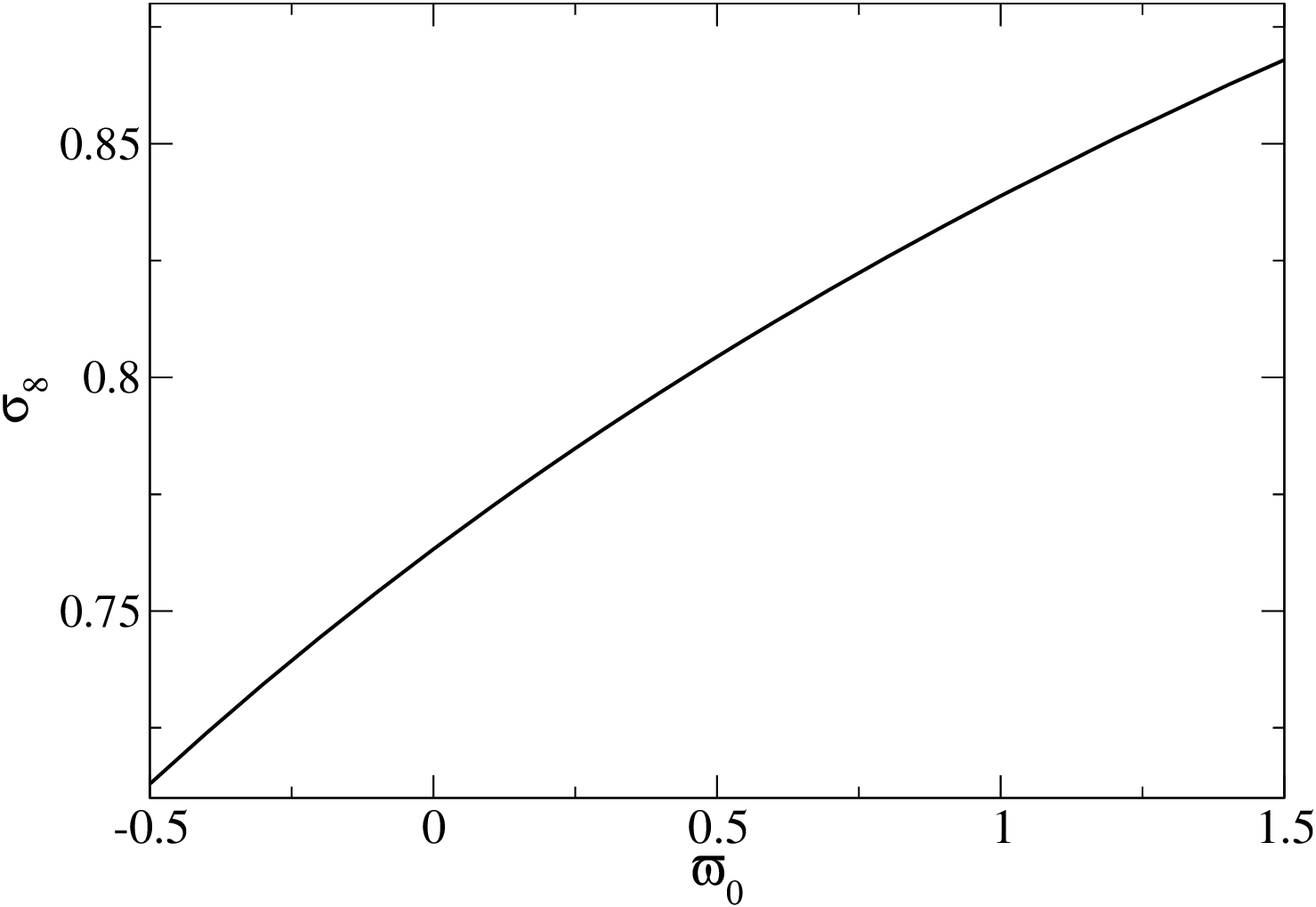}
\caption{The values of $\sigma_8$ resulting from the normalization to the WMAP
3-year anisotropy power spectrum, are shown as a function of $\varpi_0$. All other
cosmological parameters are set to match the WMAP3 ML model.}
\label{sigmafig}%
\end{figure}

The lensing convergence depends on the power spectrum of metric perturbations $\phi+\psi$. 
At linear scales, the power spectrum can be expressed through the power spectrum of density
perturbations $\delta$ by making use of the growth function
$D_\varpi=\delta(z)/\delta(z=0)$ to account for the  redshift growth of perturbations and
the Poisson equation.  Using the relation between $\phi$ and $\psi$, we can write 
\begin{eqnarray}
P_{\phi+\psi}(k,z)&=&\frac{9}{4}\Omega_{m,0}^2\left(\frac{H_0}{ck}\right)^4 \left(\frac{D_\varpi(z)}{a(z)}\right)^2\left[\frac{2+\varpi(z)}{2}\right]^2\nonumber \\
&&\quad \quad \times  P_{\delta\delta}(k,z=0) \, ,
\end{eqnarray}
where $P_{\delta\delta}$ is the power spectrum of density perturbations today. Because the
length scales probed by the weak lensing measurements extend into the non-linear regime,
this description must be extended in order to compute the non-linear matter power spectrum.
Since we do not have a complete description of non-linear evolution under modified gravity,
here we employ the same technique as used for the usual GR predictions of lensing
statistics. We use  the fitting function from Peacock and Dodds \cite{Peacock:1996ci} (PD)
to get $P^\text{nonlin}_{\delta \delta}$  from the linear power spectrum calculated with
our modified CMBfast code. Note that PD's factor $g_{PD}(\Omega,a)=\delta(a)/a$ is meant to
compare the growth history to that of an $\Omega_m=1$ model \cite{Carroll:1991mt}, so
$\delta(a)$ normalized at an early redshift is used. Specifically the density contrast is
set to $\delta(a=0.01)=0.01$ for all $\varpi_0$ models.  The linear power spectrum
amplitude is set by the normalization of the CMB anisotropy spectrum to the WMAP 3 year
data. In Figure~\ref{sigmafig} we highlight the relation between $\sigma_8$ as a function
of $\varpi_0$ and note that  an increase in $\varpi_0$ has a similar effect of enhancing 
weak lensing as does an increase in $\sigma_8$ in standard $\Lambda$CDM cosmologies. For
$\varpi_0=0$, the GR case, we recover the WMAP3 ML preferred value of $\sigma_8=0.76$
\cite{Spergel:2006hy}.

\begin{figure}[ht]
\includegraphics[scale=0.35]{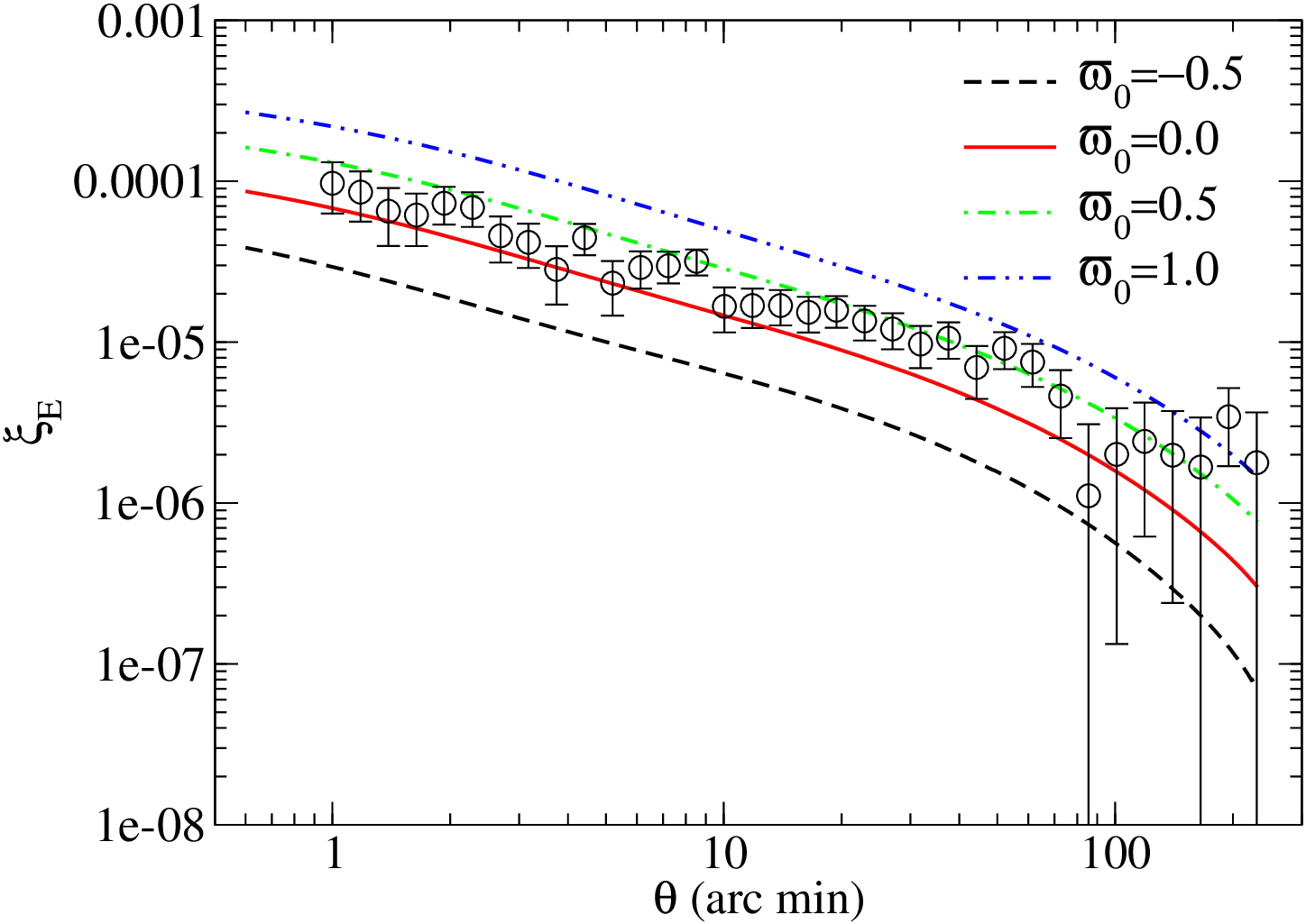} 
\caption{Model predictions of $\xi_E$ for different values of $\varpi_0$.   Data is taken
from Table B.1 of \cite{Fu:2007qq}. The background cosmology is the WMAP3 ML. The power
spectrum amplitude is determined by the normalization to WMAP.}
\label{correlationplot}%
\end{figure}

From the convergence spectrum, we can then find the modes of the correlation function
\cite{Stebbins:1996wx,Benjamin:2007ys,Munshi:2006fn,Jarvis:2005ck}
\begin{eqnarray}
\xi_{+}(\theta)&=&\frac{1}{2\pi}\int_0^\infty dk k P_\kappa(k)J_0(k\theta)\cr
\xi_{-}(\theta)&=&\frac{1}{2\pi}\int_0^\infty dk k P_\kappa(k)J_4(k\theta)\cr
\xi^\prime(\theta)&=&\xi_{-}(\theta)
+4\int_0^\infty \frac{d\theta^\prime}{\theta^\prime}\xi_{-}(\theta^\prime)
-12\theta^2\int_0^\infty \frac{d\theta^\prime}{\theta^{\prime 3}}
\xi_{-}(\theta^\prime)\cr
\xi_E(\theta)&=&\frac{\xi_{+}(\theta)+\xi^\prime(\theta)}{2} 
\end{eqnarray}
where $J_i(l)$ are Bessel functions of the first kind.  Figure \ref{correlationplot} plots
$\xi_E(\theta)$ for different $\varpi$ models with the WMAP3 ML cosmology. Varying
$\varpi_0$ results in a multiplicative shift in $\xi_E$.
 
\begin{figure}[ht]
\includegraphics[scale=0.35]{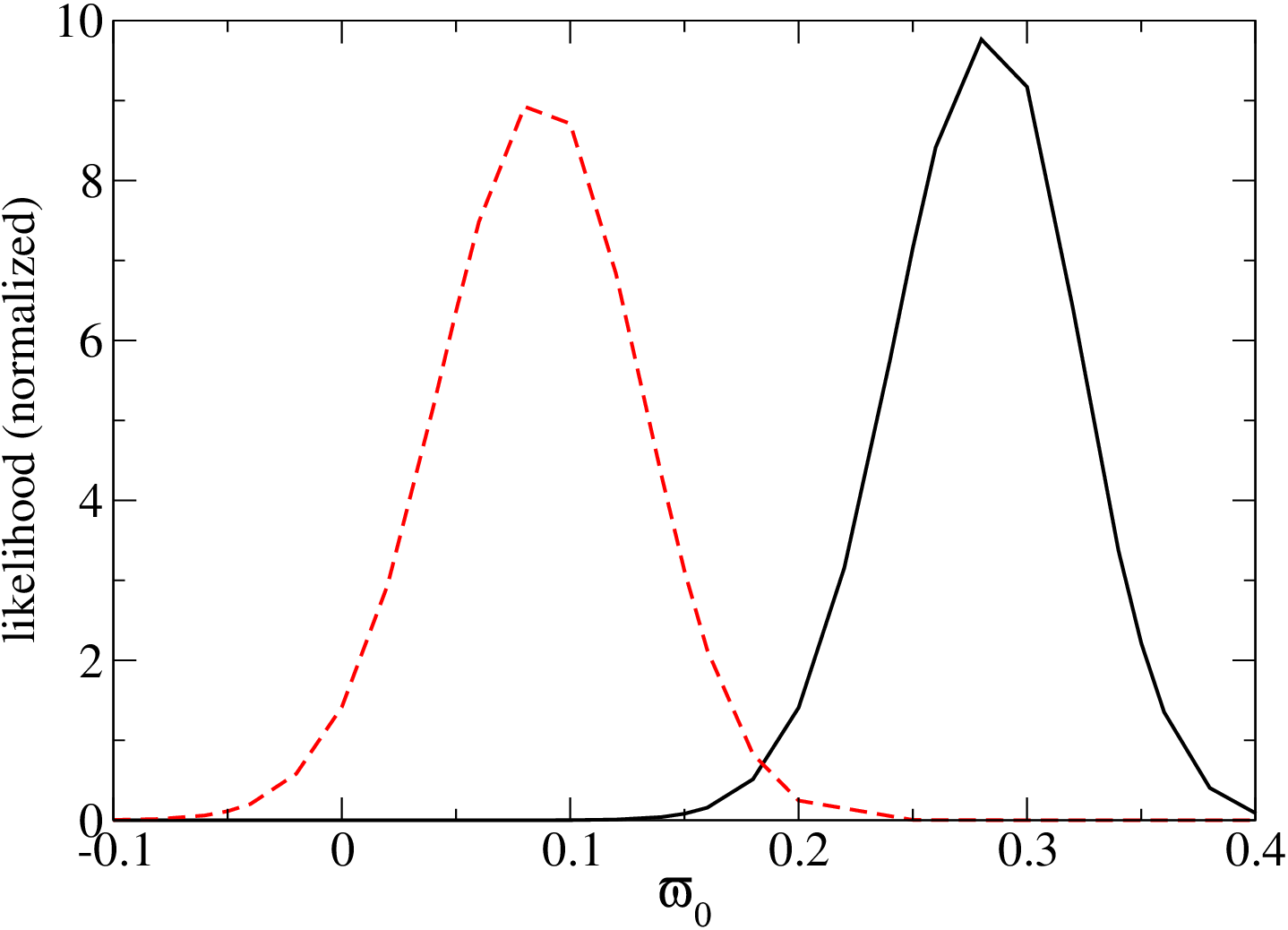}
\caption{The likelihood of different $\varpi_0$ models  according to the $\xi_E$ data
collected by the Canada-France-Hawaii Telescope Legacy Survey and presented in Table B.1 of
Ref.~\cite{Fu:2007qq} is shown with parameters set by the WMAP3 ML  model (solid line) as
well as the WMAP3+SNGold ML model (dashed line).}
\label{WLlikelihood}%
\end{figure}

Figure \ref{WLlikelihood} plots the likelihood derived from the weak lensing shear
correlation function as measured by the Canada-France-Hawaii Telescope Legacy Survey
\cite{Hoekstra:2005cs} and presented in Table B.1 of Fu {\it et al.} \cite{Fu:2007qq}.  The
source distribution $p(\chi)$ is taken from equation (14) and Table 1 of
Ref.~\cite{Fu:2007qq}. Unlike the CMB anisotropy spectra, the weak lensing shear
correlations functions do not ``bounce" with $\varpi_0$, so the resulting likelihood
function, shown in Figure~\ref{WLlikelihood}, is narrower and peaks at $\varpi_0\approx
0.3$. This result excludes $\varpi_0=0$, but is roughly consistent with the primary peak in
the CMB likelihood, shown in Figure~\ref{wmaplikelihood}.

The exclusion of $\varpi_0=0$ based on CMB and weak lensing data in the WMAP3 ML model is
not significant at this stage. Only a single parameter in a multidimensional parameter
space has been searched. Indeed, a positive value of $\varpi_0$ is expected since we have
normalized the underlying density power spectrum to the WMAP3 preferred value of
$\sigma_8$, whereas analyses of the weak lensing data have found a preferred
normalization  value with $\sigma_8\sim 0.8$, which is slightly higher than the CMB value. The
disagreement between these two values is now resolved by a non-zero value for $\varpi_0$. 
While such a difference between CMB and low-redshift matter perturbations is generally the
very type of behavior we should look for in modified gravitation scenarios, the fact that
we do not perform a joint combined analysis of CMB and weak lensing data precludes us from
making a strong statement. 

We expect a degeneracy in the effect of $\varpi_0$ and $\Omega_m h^2$ on the weak lensing
predictions, as both parameters control the growth rate of fluctuations. Specifically,
increasing $\varpi_0$ enhances fluctuation growth, as seen in Figure~\ref{deltaratio}, just
as does increasing the abundance of clustering matter. By raising $\Omega_m h^2$ the
likelihood is expected to peak at lower values of $\varpi_0$. This is precisely illustrated
in Figure~\ref{WLlikelihood}, wherein the weak lensing likelihood for $\varpi_0$ is shown
for cosmological models with parameters set by the WMAP3+SNGold ML model, which has a
slightly higher matter density. This new likelihood is indeed consistent with lower values
of $\varpi_0$, including zero.

%%%%%%%%%%%%%%%%%%%%%%%%%%%%%%%%%%%%%%%%%%%%%%%%%%%%%%%%%%%%%%%%%%%%%%%%%%%%%%%%%%%%%%%%%%%%
\section{CMB and LSS Cross-Correlations}
\label{iswsection}

A deviation from GR leaves an imprint on the cross-correlation between the CMB and
large-scale structure. In order to identify such a signal, we define the two-point angular
cross-correlation between the temperature ISW anisotropy and the dark matter fluctuation as
\cite{Crittenden:1995ak,Cooray:2001ab,Afshordi:2004kz,Corasaniti:2005pq}
\be
C^X(\theta)=\langle\Delta_{ISW}(\hat{\gamma_1})\delta_{LSS}(\hat{\gamma_2})\rangle,
\label{cross}
\ee
where the angular brackets denote the average over the ensemble and $\theta=\vert
\hat{\gamma_1}-\hat{\gamma_2}\vert$. For computational purposes it is convenient to
decompose  $C^X(\theta)$ into a Legendre series such that,
\be
C^X(\theta)=\sum_{l=2}^{\infty}\frac{2l+1}{4\pi}C_l^{X}P_l(\cos(\theta),
\label{cxpl}
\ee
where $P_l(\cos\theta)$ are the Legendre polynomials and $C_l^{X}$  is the
cross-correlation power spectrum given by 
\cite{Cooray:2001ab,Afshordi:2004kz,Garriga:2003nm,Pogosian:2004wa}
\be
C_l^X=4\pi\frac{9}{25}\int \frac{dk}{k}\Delta_{\mathcal{R}}^2 I^{ISW}_l(k)I^{LSS}_l(k),
\ee 
where $\Delta_{\mathcal{R}}^2$ is the primordial power spectrum. The integrand functions
$I^{ISW}_l(k)$ and $I^{LSS}_l(k)$ are defined respectively as
\begin{eqnarray}
I^{ISW}_l(k)&=&-\int e^{-\kappa(z)} \frac{d((2+\varpi)\phi_k)}{dz} j_l[k r(z)] dz\\
I^{LSS}_l(k)&=&b\int  \Phi(z) \delta^k(z)j_l[k r(z)] dz,\label{growth}
\end{eqnarray}
where $\phi_k$ and $\delta^k$ are the Fourier components of the gravitational potential
and matter perturbation, respectively; $\Phi$ is the galaxy survey selection function;
$j_l[k r(z)]$ are the spherical Bessel functions; $r(z)$ is the comoving distance at
redshift $z$ and $\kappa(\tau)=\int_\tau^{\tau_0}\dot{\kappa}(\tau) d\tau$ is the total
optical depth at time $\tau$.

\begin{figure}[b]
\includegraphics[scale=0.35]{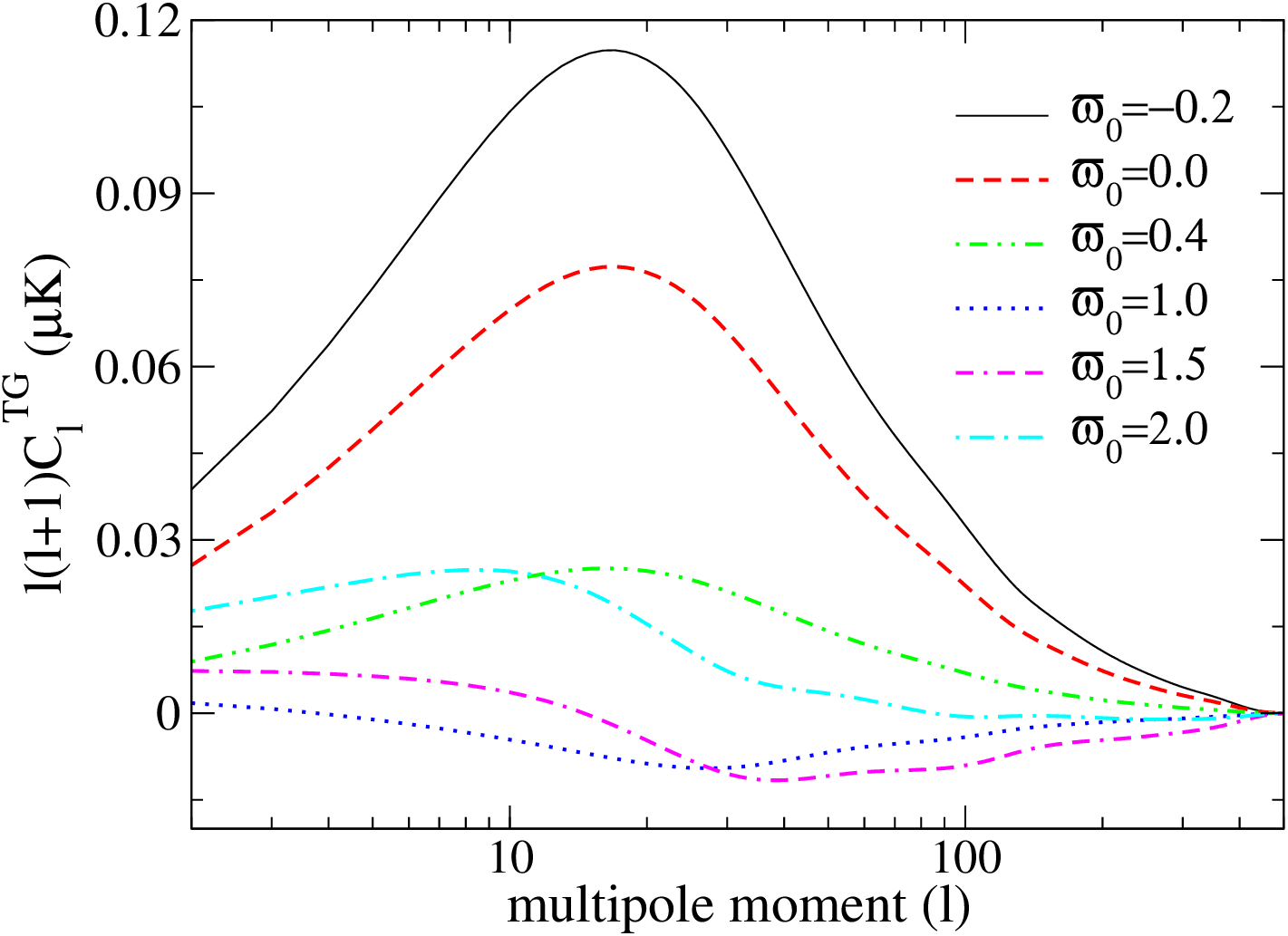}
\caption{The cross-correlation angular power spectrum between CMB temperature (ISW) and galaxy
distribution is shown as a function of $\varpi_0$. A value of $0.5<\varpi_0<1.5$ changes the sign of
the ISW, resulting in a negative cross-correlation, in conflict with observations.}
\label{iswfigure}
\end{figure}

A change in $\varpi$ could therefore change not only the value of $I^{ISW}_l$ but also its
sign with respect to $I^{LSS}_l$.  Direct measurements of the cross-power spectrum $C_l^X$ 
are more robust for likelihood parameter estimation since these data would be less
correlated than measurements of $C^X(\theta)$.  Therefore, the cross-power spectrum $C_l^X$
is computed for different values of $\varpi_0$ assuming a galaxy survey with a selection
function as
\begin{equation}
\Phi(z)\sim z^2 \exp{[-(z/\bar{z})^{1.5}]}
\end{equation} 
where $\bar{z}$, the median redshift of the survey, is  $\bar{z}=0.25$. In recent years,
the WMAP temperature anisotropy maps have been cross-correlated with several surveys of
Large Scale Structure (LSS) distributions and a positive correlation signal has been
detected
\cite{Fosalba:2003ge,Boughn:2003yz,Scranton:2003in,Nolta:2003uy,Afshordi:2003xu,giannantonio06}.
As seen in Figure~\ref{iswfigure}, values of  $\varpi_0 > 0.5$ would result in
anti-correlation and a negative angular spectrum, in disagreement with current
observations at the $\sim 2 \sigma$ level.  

\begin{figure}[t]
\includegraphics[scale=0.35]{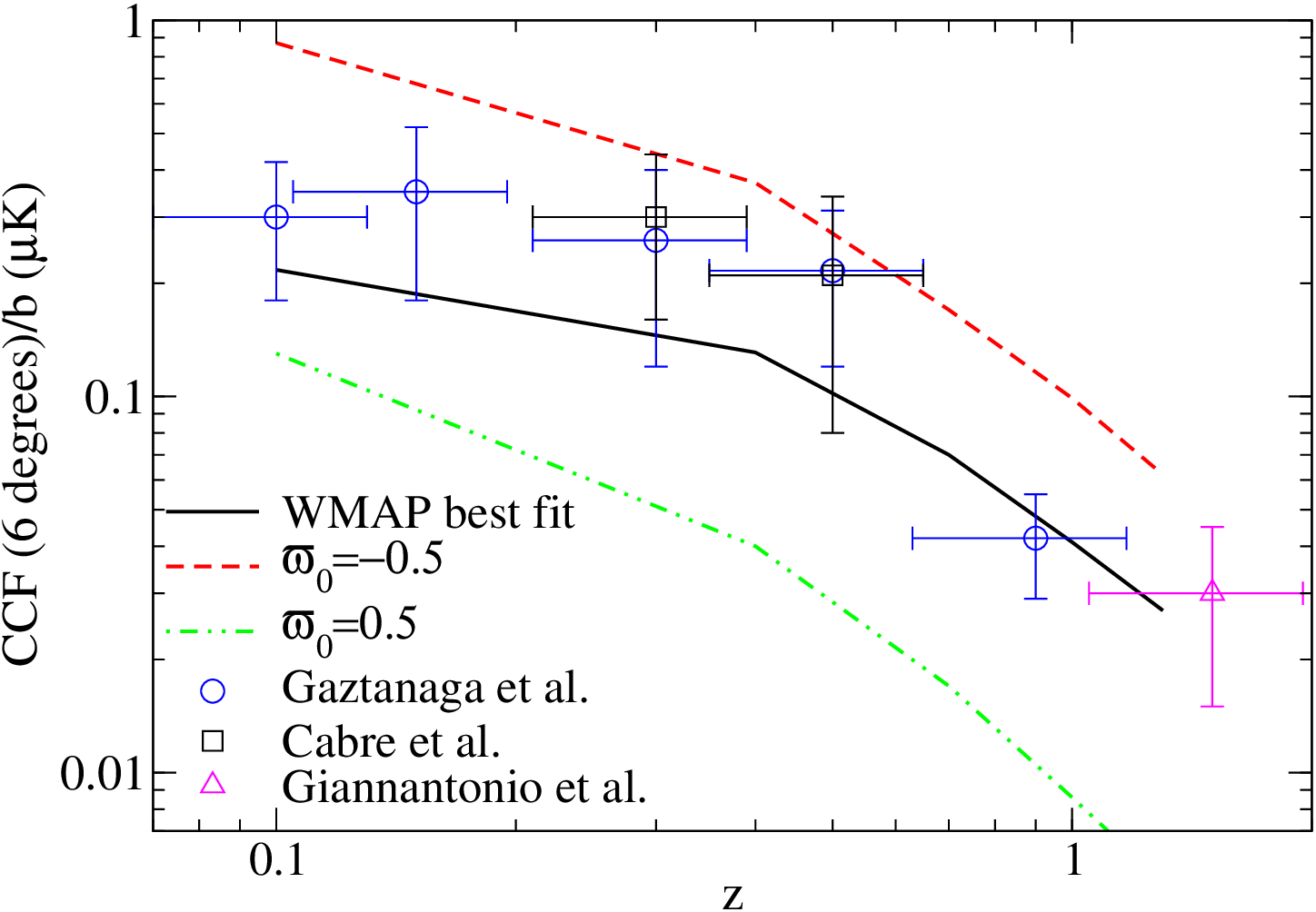}
\caption{The cross-correlation amplitude at $\theta=6$ degrees is shown as a function of
redshift for different samples of LSS and WMAP data. The curves show the expected
correlation with $\varpi_0$ varied. The data plotted here comes from the compilation in
Ref.~\cite{giannantonio06}, making also use of data from Gaztanaga {\it et
al.}~\cite{Gaztanaga:2004sk} (circles), Cabre {\it et al.}~\cite{Cabre:2006qm} (squares),
and Giannantonio {\it et al.}~\cite{giannantonio06} (triangle).}
\label{iswdata}
\end{figure}

\begin{figure}[b]
\includegraphics[scale=0.35]{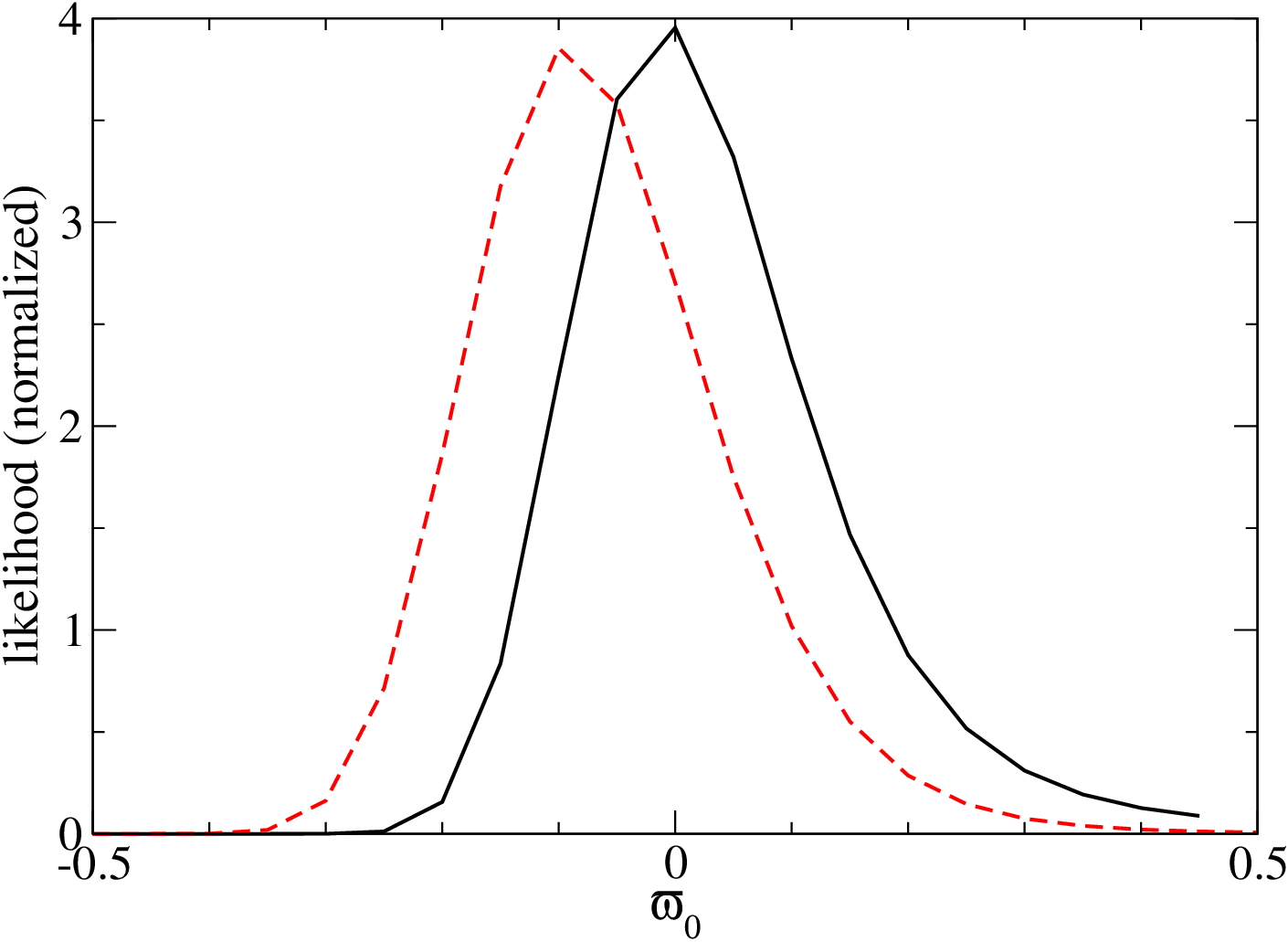}
\caption{The likelihood of different $\varpi_0$ models  using current ISW-galaxy cross
correlation data is shown. All other cosmological parameters are fixed to the WMAP3 ML
model (solid line) or the WMAP3+SNGold ML model (dashed line).}
\label{iswlike}
\end{figure}

In order to obtain a more quantitative result, the ISW-Galaxy theoretical correlation
function for different values of $\varpi$ is compared with the current data as reported in
Ref.~\cite{giannantonio06} under the assumption of the WMAP ML model. The data and some model
curves under modified gravity are shown in Figure~\ref{iswdata}. The likelihood is reported
in Figure~\ref{iswlike}, consistent with no indications for modified gravity. This
constraint should be considered with caution, since the full parameter space has not been
explored and the ISW signal may certainly be highly sensitive to changes in parameters such
as the matter density and the Hubble constant. 

Qualitatively, the ISW-Galaxy theoretical correlation may be understood by considering the
path of a CMB photon through a region on the sky containing an abundance of galaxies. As
the photon passes through the collective gravitational potential of the clustered galaxies,
the decay of the gravitational potential in a $\Lambda$CDM cosmology with $\varpi_0=0$
leads to the blueshifting of the photon. A negative value of $\varpi_0$ suppresses the
clustering and causes the photon to be further blueshifted as it climbs out of the
weakening gravitational potential. The coincidence of the hot spot in the CMB resulting
from the blueshifted photons, with the abundance of galaxies on the sky explains a
positive cross-correlation.  Likewise, a large $\varpi_0$ enhances clustering and reduces
the blueshifting as the photon climbs out of the growing gravitational potential. In
extreme cases, the enhanced clustering causes the photon to be redshifted, thereby leading
to a negative cross-correlation.  The degeneracy between $\varpi_0$ and $\Omega_m h^2$
suggests that a preference for $\varpi_0 < 0$ seen in Figure~\ref{iswlike}, will be enhanced
by increasing $\Omega_m h^2$: the slower decay of the gravitational potential ($\Gamma \to
1$ in equation \ref{iswanalysis}) due to the increased matter density is compensated by the
reduced clustering with larger $\varpi_0$. Figure~\ref{iswlike} also shows the fit to
current ISW-galaxy cross correlation data using the WMAP3+SNGold ML model parameters, for
which the matter density is slightly higher than the WMAP3 ML model. Indeed, the peak of
the likelihood shifts to even more negative values, as expected.

%%%%%%%%%%%%%%%%%%%%%%%%%%%%%%%%%%%%%%%%%%%%%%%%%%%%%%%%%%%%%%%%%%%%%%%%%%%%%%%%%%%%%%%%%%%%
\section{Synthesis}
\label{secsynth}

\begin{figure}[ht]
\includegraphics[scale=0.35]{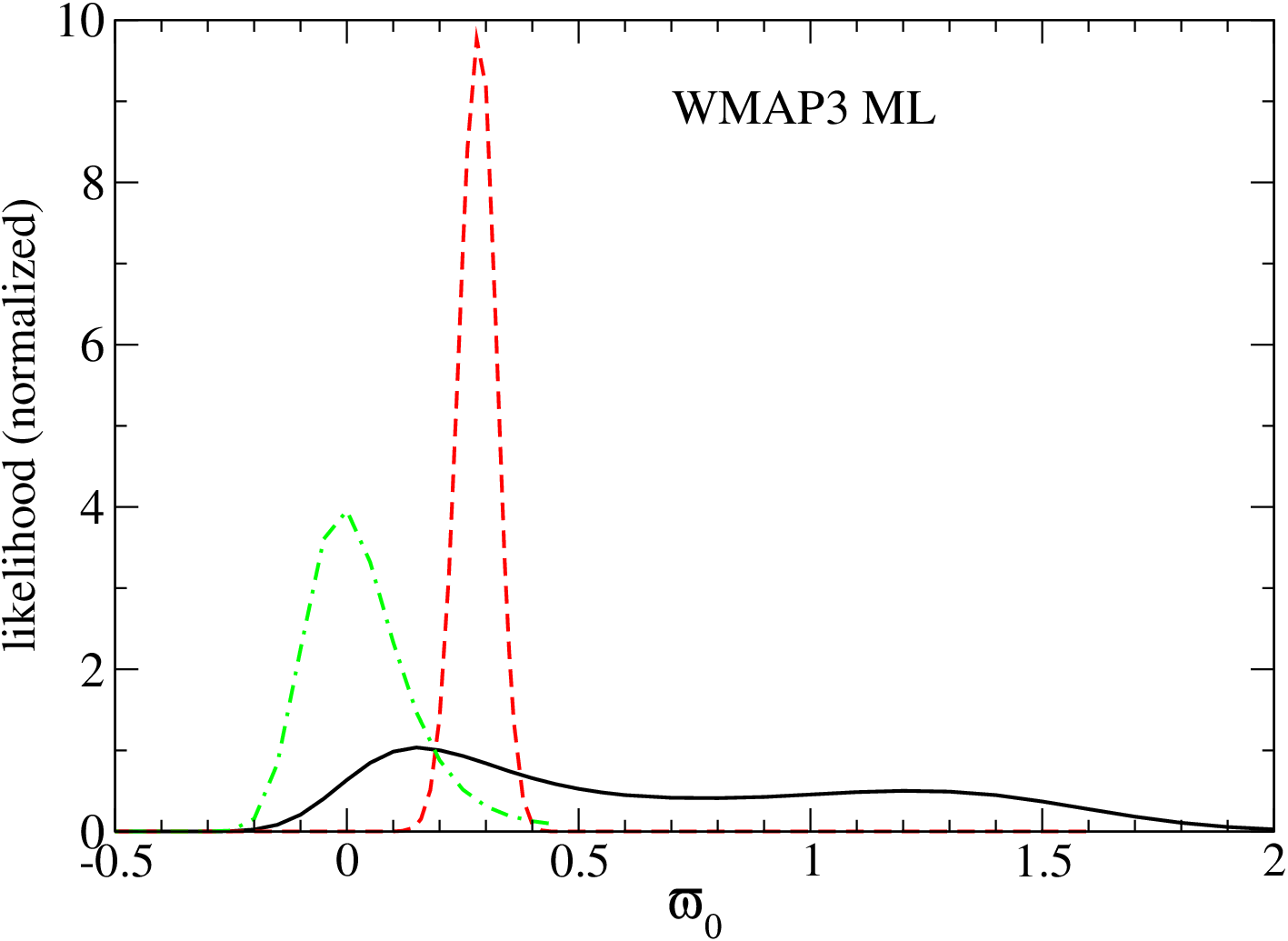}
\caption{The likelihood distributions for $\varpi_0$, based on the CMB (solid curve), weak
lensing (dashed), and ISW-galaxy cross correlation (dot-dashed) using the WMAP3 ML
model parameters is shown.}
\label{jointlikelihood}%
\end{figure}

\begin{figure}[ht]
\includegraphics[scale=0.35]{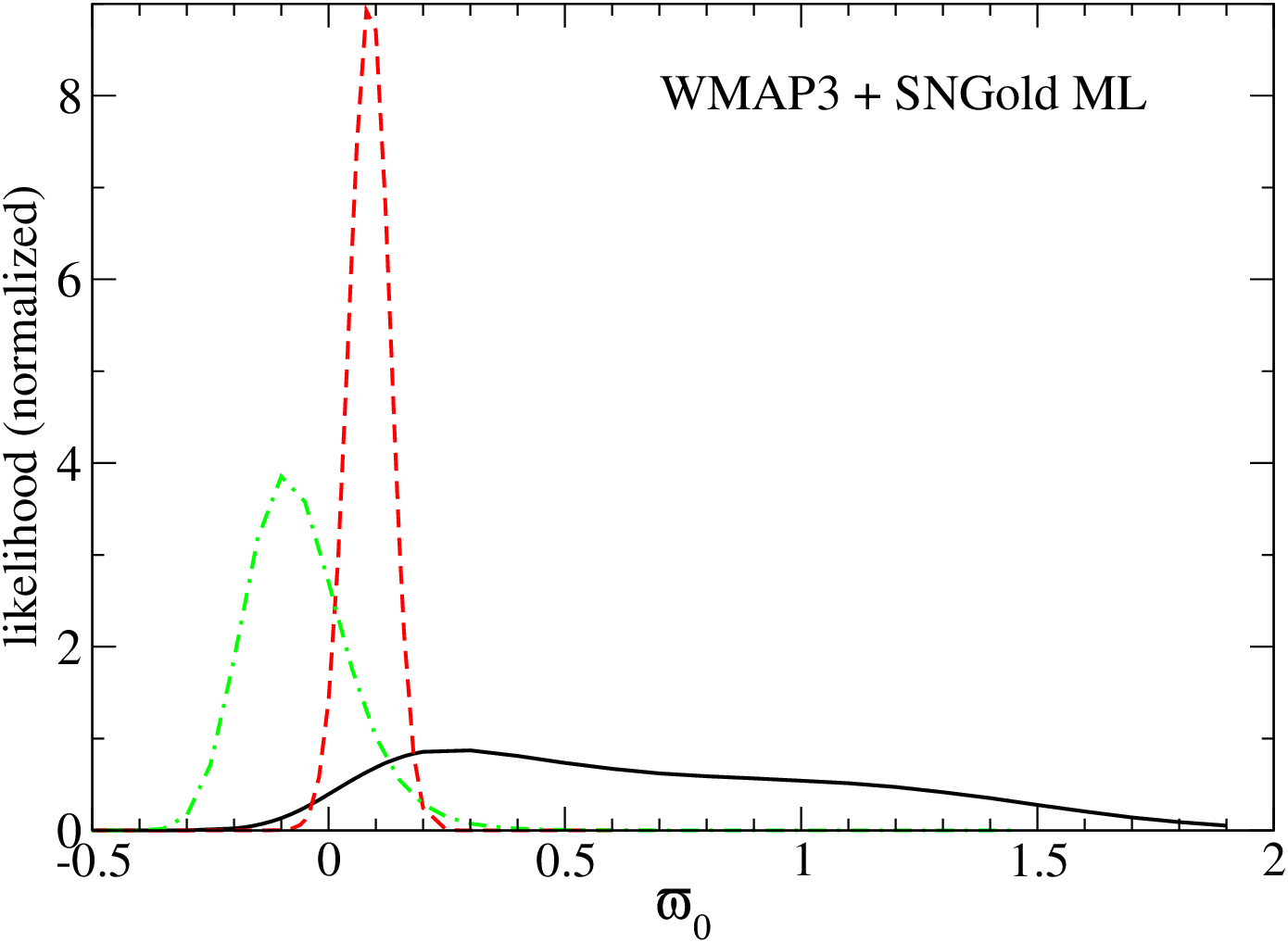}
\caption{The likelihood distributions for $\varpi_0$, based on the CMB (solid curve), weak
lensing (dashed), and ISW-galaxy cross correlation (dot-dashed) using the
WMAP3+SNGold ML model parameters is shown.}
\label{jointlikelihoodSN}%
\end{figure}

We have examined the consequences of the proposed PPF model of gravitational slip for
predictions of the cosmic microwave background anisotropy, the growth of large scale
structure, weak lensing, and the ISW-galaxy cross-correlation. The likelihood distributions
in $\varpi_0$ using the WMAP3 ML model parameters are summarized in
Figures~\ref{jointlikelihood}, and using the WMAP3+SNGold ML model parameters in
Figure~\ref{jointlikelihoodSN}. The weak lensing is clearly the most sensitive to
$\varpi_0$, followed by the ISW-galaxy cross-correlation, and then the CMB. 

In the case of the WMAP3 ML model parameters, shown in Figure~\ref{jointlikelihood}, the
overlap between the three distribution functions is strongest near $\varpi_0 \approx 0.2$.
However, there is no concordance due to the tension between the values $\varpi_0 > 0$
indicated by the CMB and weak lensing data, and the values centered on zero for the
ISW-galaxy cross-correlation. This tension appears to be related to the disagreement in the
best-fit $\sigma_8$ values derived from measurements of the CMB and large scale structure.
In fact, the additional degree of freedom introduced by $\varpi_0$ seems to be justified by
the sharp peak in the weak lensing distribution.

To find concordance among the three observational constraints, we increased the matter
density. As we argued, this should have the effect of shifting the weak lensing
distribution towards lower values of $\varpi_0$, with the primary concern to determine
whether $\varpi_0=0$ is allowed. Increasing the matter density also has the effect of moving
the CMB distribution towards higher values, away from $\varpi_0=0$, and moves the
ISW-galaxy cross-correlation towards more negative values, also away from $\varpi_0=0$.
Using the WMAP3+SNGold ML model parameters, as shown in Figure~\ref{jointlikelihoodSN}, the
overlap is improved. The tension between weak lensing and ISW is somewhat relaxed, with a
joint likelihood that allows $\varpi_0 = 0$. Whether $\varpi_0=0$ is preferred remains to
be demonstrated. A full, multi-dimensional parameter space analysis is planned for a
subsequent paper.

%%%%%%%%%%%%%%%%%%%%%%%%%%%%%%%%%%%%%%%%%%%%%%%%%%%%%%%%%%%%%%%%%%%%%%%%%%%%%%%%%%%%%%%%%%%%
\section{Discussion}
\label{secdiscuss}
  
A PPF formalism for modified gravity has been proposed to describe possible departures from
cosmological predictions under GR. The model consists of a background cosmology, a
parameterized relationship between the gravitational potentials, and an implementation of
new evolution equations in the absence of the perturbed Einstein's equations. For
simplicity we have focused on the case in which the background cosmology  evolves as a
standard, $\Lambda$CDM universe. The gravitational slip is chosen to evolve in proportion
to the dominance of an effective dark energy density over matter. The constant of
proportionality introduces a new parameter, $\varpi_0$. The new evolution equations are
summarized by equations (\ref{alphadot}-\ref{hdot}).

We have made a comparison between our proposed PPF model of modified gravitation and other
models that have recently appeared in the literature. Our implementation satisfies the
consistency relation proposed by Bertschinger \cite{Bertschinger:2006aw}. Furthermore, the
equivalence with the model of BZ has been demonstrated, after accounting for the
correspondence between $\varpi(z)$ and $\gamma_{\rm BZ}(a)$, using analytical calculations
and numerical results.  

We remark that many different parameterizations for the departure from GR on cosmological
scales have been introduced in the literature. This variety may be useful, as no single
model of non-Einstein gravitation stands out. However, it would also be useful if the data
are compared within a single parameterization, similar to the post-Newtonian parameter
$\gamma_{\rm PPN}$ that is used to measure the amount of spacetime curvature per unit mass
in tests of gravitation in the solar system. In that regard, our parameter corresponds to
$\varpi \approx 1-\gamma_{PPN}$ in the limit of weak departures from GR. When $\varpi$ is
positive, then the gravitational potential determined by geodesic motion is greater than
the potential inferred from the distribution of matter via the Poisson equation.

An alternative view of the gravitational slip is that it effectively introduces a new
source for the perturbed off-diagonal space-space Einstein equation. If we infer the
existence of some new density perturbations, defined so as to balance the perturbed
time-time Einstein equation, then the new source takes the simple form $\frac{1}{3}\varpi\,
\delta\rho_{\rm tot}$.

We have corrected an error in CCM, and proceeded to examine the cosmological consequences
of $\varpi_0\neq 0$. We hope our study clarifies that: (a) there is a tension in current
data that may be suggestive of PPF modifications of GR; (b) different parameterizations
produce the same results at the end as long as certain consistency relations are satisfied;
and (c) it would be useful to establish one or two (and same) parameters from future data
to finely test the departures from GR at cosmological scales.

%%%%%%%%%%%%%%%%%%%%%%%%%%%%%%%%%%%%%%%%%%%%%%%%%%%%%%%%%%%%%%%%%%%%%%%%%%%%%%%%%%%%%%%%%%%%
\appendix

\section{Equivalence between models of PPF gravitational slip}
\label{zetasection}

Bertschinger and Zukin \cite{Bertschinger:2008zb} propose that $\phi$ evolves according to
equation (\ref{phidotbz}) after $z=30$, with constant curvature perturbations, $\zeta$. To
show that their model is consistent with $\varpi\Lambda$CDM, their evolution equation is
recast as
\begin{eqnarray}
\zeta&=&\phi+\frac{\dot{\phi}+\mathcal{H}\psi}{\mathcal{H}(1-\dot{\mathcal{H}})}\nonumber\\
&=&\eta+\frac{2}{3}\frac{\dot{\eta}}{\mathcal{H}(1+w)}\label{zetadef}
\end{eqnarray}
where the translation from conformal-Newtonian (longitudinal) gauge to synchronous gauge,
\begin{eqnarray}
\phi&=&\eta-\mathcal{H}\alpha\label{phidef}\\
\psi&=&\dot{\alpha}+\mathcal{H}\alpha\, ,\label{psidef}
\end{eqnarray}
is used in the second equality. Also, in a matter- and $\Lambda$-filled universe,
$\mathcal{H}(1- \dot{\mathcal{H}}/\mathcal{H}^2) =\frac{3}{2}\mathcal{H}(1+w)$ where $w$ is
the background equation of state. 

For consistency with the $\varpi\Lambda$CDM model, the evolution of $\eta,\,\dot\eta$
according to equations (\ref{alphadot}-\ref{hdot}) must agree that $\zeta$ as given in
equation (\ref{zetadef}) is a constant. Hence, assuming the zero-i perturbed Einstein
equation (equation 21b of \cite{Ma:1995ey}) then
\begin{eqnarray}
\zeta&=&\eta+\left(\frac{2}{3}\right)\frac{4\pi G a^2 (\rho+p)\theta}{k^2\mathcal{H}(1+w)}\nonumber\\
&=&\eta+\frac{\mathcal{H}\theta}{k^2}\nonumber
\end{eqnarray}
where the background Friedmann equation is used to get the second equality. Taking a 
derivative with respect to conformal time,
\begin{equation}
\dot{\zeta}=\dot{\eta}+\frac{\dot{\mathcal{H}}\theta}{k^2}
+\frac{\mathcal{H}\dot{\theta}}{k^2}\, .
\end{equation}
Again using the zero-i Einstein equation, and the equation for $\dot{\theta}$ derived from
stress-energy conservation (equation 29 of \cite{Ma:1995ey}),  
\begin{equation}
\dot{\theta}=-\mathcal{H}(1-3w)\theta-\frac{\dot{w}}{1+w}\theta
+\frac{\delta p/\delta\rho}{1+w}k^2\delta-k^2\sigma\, ,
\end{equation}
then the time derivative of the curvature perturbation is
\begin{equation}
\dot{\zeta}=\mathcal{H}(\frac{\delta p}{\delta\rho}\frac{\delta}{1+w}-\sigma)\, .
\end{equation}
Since the dominant perturbations at late times ($z< 30$) are due to baryonic and dark
matter that have negligible pressure and shear perturbations, then $\dot\zeta \approx 0$.
Therefore, $\varpi\Lambda$CDM is consistent with the evolution equation (\ref{zetadef})
with constant $\zeta$.

%%%%%%%%%%%%%%%%%%%%%%%%%%%%%%%%%%%%%%%%%%%%%%%%%%%%%%%%%%%%%%%%%%%%%%%%%%%%%%%%%%%%%%%%%%%%

%%%%%%%%%%%%%%%%%%%%%%%%%%%%%%%%%%%%%%%%%%%%%%%%%%%%%%%%%%%%%%%%%%%%%%%%%%%%%%%%%%%%%%%%%%%%
\vspace{0.5cm}
\acknowledgments

It is a pleasure to thank Tommmaso Giannantonio for help. SFD and RRC were supported in
part by NSF CAREER AST-0349213. AC was supported by NSF CAREER AST-0645427. AC, SFD, and
RRC thank the Theoretical Astrophysics Group at Caltech for hospitality while this research
was conducted. AM research has been supported by ASI contract I/016/07/0 ``COFIS".

\vfill

%%%%%%%%%%%%%%%%%%%%%%%%%%%%%%%%%%%%%%%%%%%%%%%%%%%%%%%%%%%%%%%%%%%%%%%%%%%%%%%%%%%%%%%%%%%%

%%%%%%%%%%%%%%%%%%%%%%%%%%%%%%%%%%%%%%%%%%%%%%%%%%%%%%%%%%%%%%%%%%%%%%%%%%%%%%%%%%%%%%%%%%%%  

\end{document}